\documentclass[preprint,prd,tightenlines,nofootinbib,superscriptaddress]{revtex4}

\usepackage{color}
\usepackage{graphicx}
\definecolor{red}{rgb}{1,0,0}
\def\lesssim{\ \hbox{\raise 2pt \hbox{$<$} \kern -13pt
                     \lower 3pt \hbox{$\sim$}}\ }
\def\greatersim{\ \hbox{\raise 2pt \hbox{$>$} \kern -13pt
                     \lower 3pt \hbox{$\sim$}}\ }
\def\cascade{{\sc Cascade}}
\def\herwig{{\sc Herwig}}
\def\pythia{{\sc Pythia}}
\def\disent{{\sc Disent}}
\def\mcatnlo{MC@NLO}
\def\prp{\perp}
\def\kt{\ensuremath{k_\prp}}
\input epsf.tex
\def\desepsf(#1 width #2){\epsfxsize=#2 \epsfbox{#1}}

\usepackage{amsmath,bm}
\begin{document}

\hspace*{12 cm} {\small DESY-08-057} 

\hspace*{12 cm} {\small OUTP-08-13-P}

\vspace*{1 cm} 

\title{Angular correlations in multi-jet final states from\\ 
k$_\perp$-dependent parton showers}
\author{F.\ Hautmann} 
\affiliation{Theoretical Physics Department, 
University of Oxford,    Oxford OX1 3NP}
\author{H.\ Jung}
\affiliation{Deutsches Elektronen Synchrotron, D-22603 Hamburg}

\begin{abstract}
We investigate parton-branching methods based on 
transverse-momentum dependent (TMD) parton distributions 
and matrix elements  
for  the  Monte Carlo simulation of 
multi-particle final states at 
high-energy colliders. 
We observe that  recently measured angular correlations in  
 ep final states with multiple hadronic jets  
probe QCD coherence effects in the space-like  branching,  
associated with finite-angle gluon radiation from partons   carrying  small 
longitudinal momenta,   
and not included in standard shower generators. 
We present Monte Carlo calculations for 
azimuthal two-jet and three-jet  distributions, for 
jet multiplicities and 
for correlations in the transverse-momentum imbalance 
between the leading 
jets.  We discuss   comparisons  with  current experimental multi-jet  data,   and 
 implications of corrections to collinear-ordered showers for 
LHC final states.   
\end{abstract}


\maketitle

\section{Introduction}
\label{sec:intro}

Hadronic final states containing multiple jets 
have been investigated at the  Tevatron and HERA colliders, and 
 will play 
a central role  in   the   
Large Hadron Collider (LHC) 
physics program. 
The interpretation of  experimental data for such  
final states relies  
  both on  perturbative  multi-jet calculations  
  (see~\cite{bernhouches} for a recent overview) and 
 on  realistic  event simulation  by 
 parton-shower Monte Carlo generators   (see  
e.g.~\cite{mlmhoche,alwalletal,heralhcproc}).

Owing to the complex 
kinematics involving multiple 
hard scales and the  large phase space opening up at very 
high  energies, multi-jet events 
are  potentially   sensitive 
to effects of QCD  initial-state  radiation 
 that depend on the finite transverse-momentum 
tail of partonic matrix elements and 
distributions. For an overview see~\cite{hj_rec}. 
In perturbative multi-jet 
calculations 
truncated to fixed order in $\alpha_s$~\cite{bernhouches},   
 finite-\kt \ contributions 
 are taken into account  partially, 
order-by-order,  through higher-loop  corrections. 
 This is generally  sufficient 
  for inclusive jet cross sections, but likely not 
  for  more  exclusive final-state observables.

On the other hand, 
standard shower Monte Carlos reconstructing exclusive events,   
 such as  
\herwig~\cite{herwref} and 
\pythia~\cite{pythref},  are based on 
collinear evolution of the initial-state jet. 
Finite-\kt \ contributions are not included, but rather  
 correspond to corrections~\cite{mw92,skewang,hef} to the 
angular or transverse-momentum ordering  implemented in the 
parton-branching   algorithms.
The theoretical framework  to take these corrections into account is based on 
using initial-state  distributions (pdfs) unintegrated in 
both longitudinal and transverse momenta~\cite{hef},   
coupled to hard  matrix elements (ME)
suitably  defined off mass shell. See e.g.~\cite{boand02} 
for discussion of the   
Monte Carlo shower implementation of the method.  
Event generators  based on \kt-dependent showers 
 of this kind 
include~\cite{junghgs,lonn,golec-mc,krauss-bfkl}. 
 These generators are not as developed as  standard 
  Monte Carlos like \herwig\  and 
\pythia. However,  they have the potential advantage of a 
more accurate treatment of  the space-like parton shower 
   at high energy.  

Collinear-based shower generators like
\herwig\ and the new \pythia\  contain the effects of color coherence 
for soft gluon emission from partons carrying longitudinal
momentum fraction $x \sim {\cal O} (1)$. However as the energy increases
and  emissions that are not collinearly ordered
 become  more important, coherence effects from
space-like partons carrying momentum fractions $x \ll 1$ set in.
 These small-x coherence effects are  not included in \herwig\ or 
  \pythia\ but are included in \kt-dependent parton showers, and characterize 
  the structure of the initial-state branching at very high energies.

This paper examines how  corrections  to space-like parton showers 
affect properties of final state jet correlations and 
associated distributions. 
We study azimuthal correlations and transverse-momentum correlations for 
multi-jet processes. We obtain numerical Monte Carlo results for 
collinear and \kt-dependent parton showers, 
and use the precise experimental data on tri-jets in ep collisions 
that have recently become available~\cite{zeus1931}. We observe 
significant   corrections 
arising from regions~\cite{hj_rec}  with
 three well-separated hard jets 
in which the partonic  lines along the decay chain
in the initial state  are not ordered in transverse momentum. 
These 
give rise to quite distinctive features in the jet angular 
 correlations. 

Besides jet final states,  the  coherence effects   from  highly 
off-shell  processes discussed in this paper 
affect a variety of different final states at high energy. 
A significant  example concerns the associated 
production of heavy flavors and heavy bosons at the LHC with two high-$p_t$ jets.  
We come back to this at the end of the paper in Sec.~\ref{sec:concl}. 

The paper is structured  as follows. We begin in Sec.~\ref{sec2} 
by describing experimental results on multi-jet correlations. 
 In Sec.~\ref{sec3} we  recall  
basic aspects on the implementation of 
transverse-momentum dependent pdfs and MEs
in parton-branching  algorithms.  
  We then compute 
angular correlations  in three-jet final states by  
\kt-dependent Monte Carlo showering. We    
 compare the results 
with  \herwig\ and with experimental data.  We consider correlations in the 
azimuthal angle between the two hardest jets, and  further 
 analyze the distribution of the third jet.  
We investigate in particular the quantitative effect of 
the finite high-\kt \  tail in the hard ME. 
   In Sec.~\ref{sec4} we present results for 
jet multiplicity distributions  
and for momentum correlations. In Sec.~\ref{sec:concl} 
we discuss  prospects for LHC final states 
and give conclusions.  
Some  details on u-pdf fits and on  time-like showering effects 
are left to  Appendix~\ref{app:ktfits} and  Appendix~\ref{sec_timelike}.

\section{Measurements of  final-state jet correlations}
\label{sec2}

In this section we recall  
experimental results from  Tevatron and 
 HERA  on angular correlations 
 in multi-jet production.   

In a multi-jet event  the  
 correlation in the azimuthal angle   $\Delta \phi$ 
 between the two hardest jets     provides  a   useful  measurement,
sensitive to how well QCD   multiple-radiation  effects are described.  
In leading order one expects two back-to-back jets; higher-order 
radiative contributions cause the  $\Delta \phi$ distribution to 
spread   out.   At the LHC, measurements of  
   $\Delta \phi$ distributions in multi-jet  events
  may  become  accessible    relatively early, and be used to 
  test the Monte Carlo description of the events. 
 
\begin{figure}
\centerline{\includegraphics[width=0.35\columnwidth]{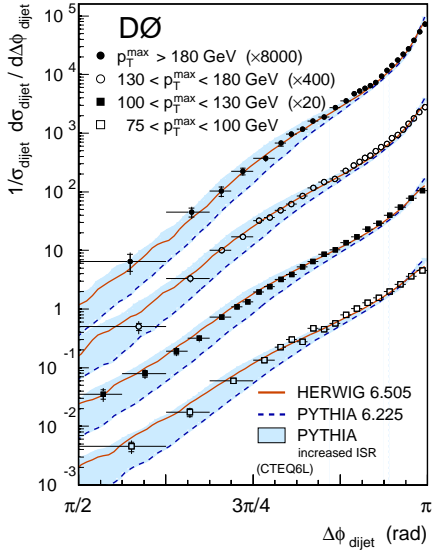}}
\caption{\label{Fig:d0az} Dijet azimuthal  correlations measured by
D0 along with the \herwig\  and  \pythia\  results~\protect\cite{d02005}.}
\end{figure}

   Fig.~\ref{Fig:d0az}~\cite{d02005}   shows 
   the  Tevatron $\Delta \phi$  measurements. The data are compared 
    with  \herwig\      and
  \pythia\               results.    The data 
    are found~\cite{d02005,albrow}   to have little  sensitivity 
    to final-state showering parameters 
and to be in contrast  very      sensitive to
     initial-state showering parameters.  
    In particular, they have been used
     for re-tuning of these parameters
     in \pythia~\cite{albrow}.   
 A  reasonably good  description  of the measurements   by
  Monte Carlo  is obtained.

On the other hand,    the
     {\small HERA}  $\Delta \phi$ 
      measurements~\cite{zeus1931,aktas_h104,hansson06}
  are not   so   well described by  the standard
   \herwig\    and   \pythia\      Monte Carlo showers    in  most
 of the data kinematic    range.  We will discuss more on this  below.
 These measurements are
characterized by the large phase space available for 
jet production and relatively 
small values of the ratio 
between the jet transverse momenta and center-of-mass energy.  
For these reasons, despite the much lower energy at HERA, 
they may be   just as  relevant   as    the Tevatron data   
for extrapolation of
initial-state showering effects to the LHC.

In the rest of this section we focus on the recent, precise 
ep measurements~\cite{zeus1931} 
of jet correlations,  
 and discuss  potential sources  of large QCD corrections.

\begin{figure}[htb]
\vspace{130mm}
\includegraphics{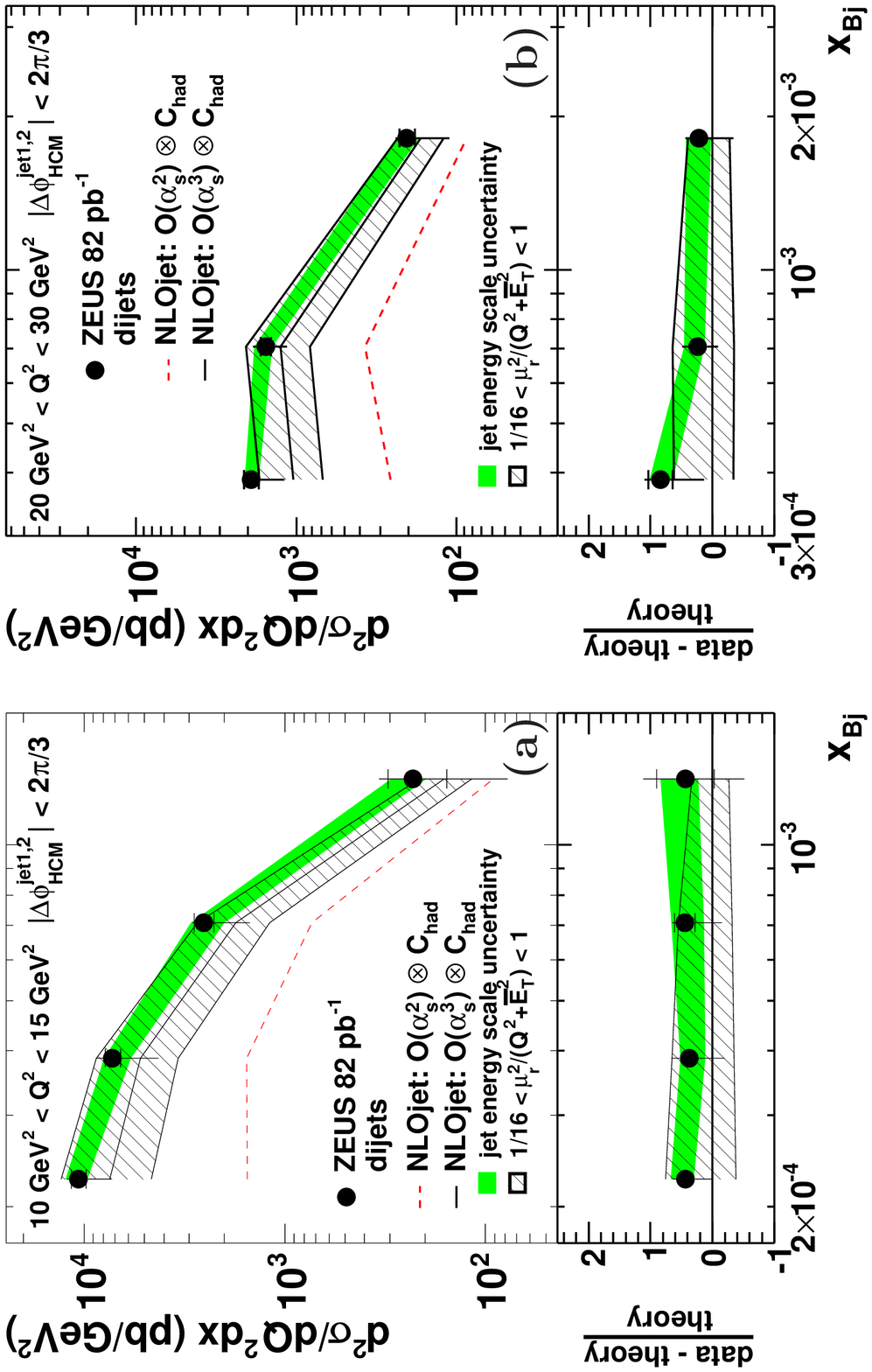}
\includegraphics{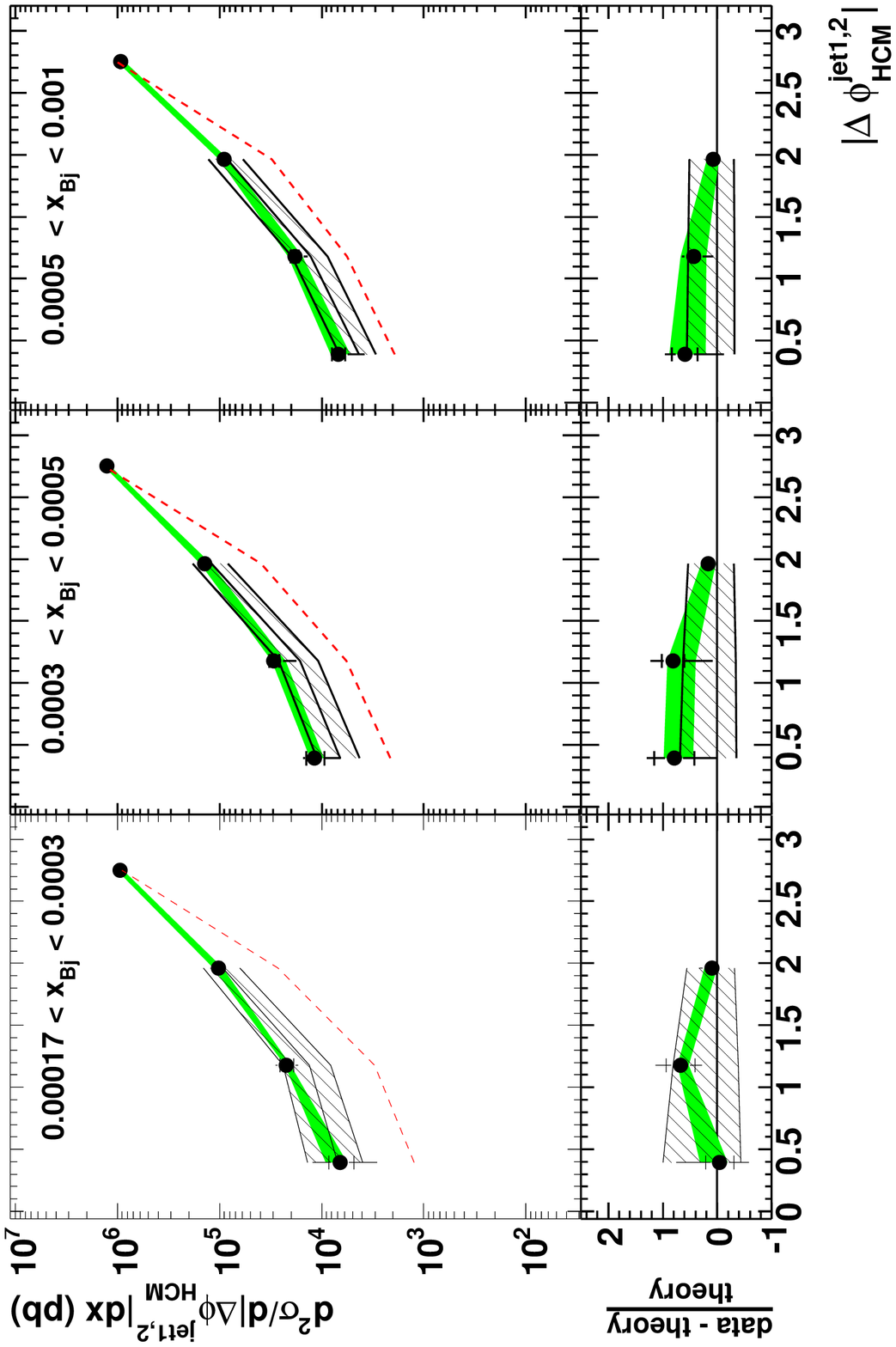}
\caption{(top) Bjorken-x  dependence and (bottom) azimuth dependence 
of  di-jet distributions at HERA as measured 
by ZEUS~\protect\cite{zeus1931}.} 
\label{fig:phizeus}
\end{figure}

In Ref.~\cite{zeus1931}  the {\small ZEUS} collaboration 
have  presented   data  for 
two-jet and three-jet production 
associated with  
\begin{equation}
\label{kinxQ}
Q^2 >  10 \ {\rm GeV}^2  \hspace*{0.3 cm} ,  \hspace*{0.6 cm} 
10^{-4} < x < 10^{-2} \hspace*{0.3 cm} , 
\end{equation} 
and performed a comparison 
  with next-to-leading-order   calculations~\cite{nagy}. 
{\small ZEUS} measured differential distributions 
 as functions of 
 jet transverse energy and 
pseudorapidity as well as  correlations in azimuthal angles 
and transverse momenta. 
The selection cuts on the jet phase space are given by 
\begin{equation}
\label{kinetaet} 
 E^{\rm{jet-1}}_{T, HCM} > 7 \ {\rm{GeV}}  \hspace*{0.3 cm} , \hspace*{0.6 cm} 
 E^{\rm{jet-2,3}}_{T, HCM} > 5 \ {\rm{GeV}}  \hspace*{0.3 cm} , 
\hspace*{0.6 cm}  - 1 < \eta_{lab} < 2.5  \hspace*{0.3 cm}    ,     
\end{equation} 
where $E_{T, HCM}$ are the jet transverse energies  
in the hadronic center-of-mass frame, and $\eta_{lab}$ are 
the jet pseudorapidities  
  in the laboratory frame. 
The overall agreement of data 
with NLO results is  
 within errors~\cite{zeus1931}. However, 
while  inclusive jet rates are 
reliably predicted by NLO perturbation theory,  
jet correlations  turn out to be affected by 
  large theoretical uncertainties 
at NLO. Results from~\cite{zeus1931} for di-jet distributions are 
reproduced in Fig.~\ref{fig:phizeus} for easier reference.

The plot at the top in Fig.~\ref{fig:phizeus} shows the 
$x$-dependence of the  di-jet distribution integrated  over 
$\Delta \phi < 2 \pi / 3$, where $\Delta \phi$ is the 
azimuthal separation 
between the two high-$E_T$ 
jets. The  plot at the bottom
shows the di-jet
distribution in $\Delta \phi$  for different bins of $x$.
We see that the variation of the 
predictions from  order-$\alpha_s^2$ to order-$\alpha_s^3$ 
is  significant.  In the  azimuthal correlation  for a given $x$ 
 bin,  the variation increases 
 with decreasing   $\Delta \phi$. 
 In the distribution integrated over  $\Delta \phi$, 
 the variation increases with decreasing $x$. 
The lowest  order, where the differential cross
section ${d\sigma} / {d\Delta \phi}$ is non-trivial, is 
 ${\cal O}(\alpha_s^2)$  and the NLO calculation is labeled with 
 ${\cal O}(\alpha_s^3)$.

Given the large difference between 
order-$\alpha_s^2$ and order-$\alpha_s^3$ results,
it seems to be questionable to estimate 
the theoretical uncertainty at NLO  from the conventional method of 
varying the renormalization/factorization scale.

Besides  angular distributions, a  behavior similar to that 
described above 
is also found in~\cite{zeus1931}  
for other associated distributions such as 
momentum correlations.\footnote{On the other 
hand,   NLO results 
are  much more stable in the case of   inclusive 
jet cross sections~\cite{zeus1931}.} 
 We will come back to this in Sec.~\ref{sec4}. 

Note that 
the  Tevatron $\Delta \phi$ distribution  in  Fig.~\ref{Fig:d0az} drops by 
two orders of magnitude  over a fairly narrow range, essentially 
 still close to the two-jet region. The measurement is dominated  essentially 
 by   leading-order processes. Not surprisingly the 
 Monte Carlos provide a good description of the data.  
 In Fig.~\ref{fig:phizeus} a comparable two order of magnitude drop  occurs 
 over the whole $\Delta \phi$ range. Much  more QCD dynamics  than leading order 
 is probed.

The stability of predictions for 
the jet observables under consideration   in Fig.~\ref{fig:phizeus}   
depends on a number of  physical effects.  
Part of these    concern the 
jet reconstruction and hadronization. 
 The {\small ZEUS}~\cite{zeus1931} and 
{\small H1}~\cite{aktas_h104,hansson06} 
 jet algorithm
has  moderate 
hadronization corrections~\cite{h101} 
and is free of nonglobal single-logarithmic  
components~\cite{nonglobjet}. 
 The kinematic cuts~\cite{zeus1931}  on the  hardest  jet 
transverse momenta are set to be 
asymmetric, so as to avoid 
 double-logarithmic contributions 
in the minimum $p_T$~\cite{dasguban}. 
Note that $Q^2 > 10$ GeV$^2$ (Eq.~(\ref{kinxQ})), 
and nonperturbative corrections affecting  the 
jet distributions at the level of inverse powers of $Q$ 
 are expected to be 
moderate.

\begin{figure}[htb]
\vspace{80mm}
\includegraphics{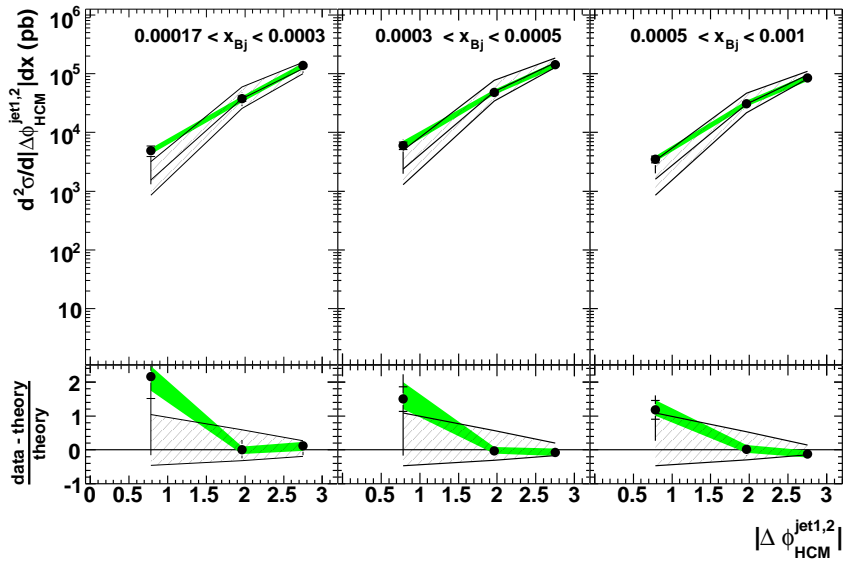}
\caption{Three-jet cross section  versus azimuthal separation between the two 
highest-$E_T$ jets as measured by ZEUS~\protect\cite{zeus1931}.} 
\label{fig:phizeus-3j}
\end{figure}

Further effects concern radiative corrections at  higher order.  
Fixed-order 
calculations beyond NLO are not within present reach 
 for multi-jet processes in $ep$ and $pp$ collisions.   
 Resummed calculations of higher-order logarithmic 
 contributions from    multiple infrared 
emissions are performed with next-to-leading accuracy in~\cite{delenda}.  
These contributions are 
  enhanced in  the region where the two high-$E_T$ 
jets are nearly back-to-back. Multiple infrared emissions 
are  also taken into account  by parton-branching methods 
 in shower Monte Carlos such as \herwig~\cite{herwref}. 
Note however that important corrections 
in Fig.~\ref{fig:phizeus}  arise  for 
decreasing $\Delta \phi$, where   
the two jets are not close to back-to-back  and 
one has effectively three  
well-separated hard jets~\cite{hj_rec}. 
The corrections increase as $x$ decreases. 
 Effects analogous to those in  Fig.~\ref{fig:phizeus} 
are  seen in the  {\small ZEUS} results  
for the three-jet cross section, shown 
in  Fig.~\ref{fig:phizeus-3j}~\cite{zeus1931}, particularly for  the 
   small-$\Delta \phi$  and small-$x$ bins.\footnote{The error band 
for the theory curves in Fig.~\ref{fig:phizeus-3j}~\cite{zeus1931}  
is obtained by varying the value of the 
renormalization scale from $( Q^2 + {\overline E}_T^2 )$  to  
$( Q^2 + {\overline E}_T^2 ) / 16$, where 
${\overline E}_T$ is the average $E_T$ of the three 
hardest jets in each event.} 
  In Sec.~\ref{sec3}  we    
analyze the  angular distribution of 
the third jet, and  find  
significant contributions 
  for small $\Delta \phi$ 
 from regions of  the space-like shower where 
the transverse momenta in  the initial-state decay chain 
  are not ordered. 
These contributions are not fully taken into account either by 
fixed-order calculations truncated to NLO or by parton showers 
implementing  collinear ordering 
 such as  \herwig\   and  \pythia.

 In the next section  we 
present the results of    computing  jets'  angular  correlations 
by parton-shower methods that include finite-\kt \  corrections to 
collinear ordering. 
 We compare these results with  the collinear-based   shower  \herwig, and with 
 experimental data.

\section{Angular  correlations from \boldmath{\kt} shower Monte Carlo}
\label{sec3}

Corrections to the collinear ordering 
in the space-like   shower  
can be incorporated in Monte Carlo  event 
generators by implementing   
transverse-momentum dependent   (TMD) 
parton distributions (unintegrated pdfs) and 
 matrix elements (ME) 
through  high-energy factorization~\cite{hef}.   
This method allows parton distributions at 
fixed \kt \ to be defined gauge-invariantly for 
small $x$.  
Basic aspects of  the 
 parton-shower implementation of the method 
are discussed in~\cite{boand02}.   
In this section we start by briefly recalling the basis for the 
introduction of unintegrated pdfs (u-pdfs) at high energy;   we comment on  
generalizations relevant for low energies and general-purpose tools; then we 
  apply  the  \kt-dependent   parton   branching to the  study of angular 
jet correlations.

\subsection{Unintegrated pdfs}
\label{sec2prime}

To characterize a 
 transverse-momentum dependent   parton 
distribution  gauge-invariantly 
over the whole phase space 
is a nontrivial  question~\cite{jcc-lc08,collins01},  currently
   at the center  of  much   
activity.  See overview in~\cite{hj_rec}.  
In the case of small $x$ a well-prescribed, gauge-invariant definition 
emerges from high-energy factorization~\cite{hef}, and has been used for studies 
of  collider processes both  by Monte Carlo (see reviews in~\cite{jeppe04,hj_rec}) 
 and by 
semi-analytic resummation  approaches (see~\cite{radcortalk1,radcortalk2}).

The diagrammatic argument for gauge invariance,  
 given in~\cite{hef}, and developed in~\cite{hef94}, is based on relating 
off-shell matrix elements with physical 
cross sections  at $x \ll 1$, and exploits   the dominance of  
single gluon polarization at high energies.\footnote{It 
is emphasized e.g. in~\cite{jeppe04,collins01} that 
a fully worked out 
 operator argument, on the other hand,  is highly desirable but is still 
missing.} 
The main reason why a natural definition for TMD 
pdfs can be constructed in the high-energy limit 
is that   one can relate directly (up to 
perturbative corrections) the cross section for a {\em physical}  
process, say, photoproduction of a heavy-quark pair, 
to an {\em unintegrated} gluon distribution, much as, in the conventional 
parton picture, one does for DIS in terms of ordinary (integrated) parton 
distributions. 
On the other hand, the difficulties 
in defining a TMD distribution over the whole phase space can 
largely be associated  with the fact that 
it is not  obvious how to determine one such 
 relation for general kinematics. 

The evolution equations obeyed by TMD distributions defined from the 
 high-energy limit are of the type of energy evolution~\cite{lip97}.  
 Factorization formulas in terms of TMD  distributions~\cite{hef}  
have corrections that are down   by logarithms of energy 
rather than powers of momentum transfer. On the other 
hand, it is important to observe that 
 this framework   allows one 
 to describe the ultraviolet region of arbitrarily high k$_\perp$   
and in particular re-obtain the structure of QCD 
logarithmic scaling violations~\cite{radcortalk1,radcortalk2,hef94}. 
  This  ultimately 
justifies the use of this approach for jet physics.       
 In particular it is the basis for  using  corresponding  
  Monte Carlo implementations~\cite{boand02,junghgs,lonn,golec-mc,krauss-bfkl,hj04}
  to treat multi-scale hard processes at the LHC.

From both 
theoretical and phenomenological viewpoints,   
it is one of the 
appealing features  of the high-energy framework for  TMD distributions 
that one can  relate its results  to a well-defined summation of 
higher-order radiative corrections. By expanding these 
results to fixed order in $\alpha_s$, one can match  the predictions thus obtained 
against perturbative calculations. This has been verified  for  a number of specific 
processes at next-to-leading order (see for instance~\cite{vannee} for heavy flavor 
production) and more recently at next-to-next-to-leading order (see for 
instance~\cite{mochetal}). 
Note that this fact also provides the basis for  
 shower algorithms implementing this framework to be combined with 
fixed-order NLO  calculations by using existing techniques for such 
matching.

Later in this section we use 
Monte Carlo implementing the high-energy definition of u-pdfs to analyze 
jet production. Before doing this, we comment briefly on open issues and 
generalizations to low energies.

\subsection{Comments on unintegrated pdfs beyond low  x}
\label{sec_lowenergy}

In the general case,   
 factorization formulas 
in terms of unintegrated parton distributions will have a considerably 
complex structure~\cite{jcc-lc08}.   Full results   are yet to be established. 
A prototypical calculation that illustrates this structure 
 is carried out in~\cite{jccfh00}, which treats,  rather 
than a general scattering  observable,  a simpler problem,  the 
electromagnetic form factor of a quark. 
This case is however sufficient to illustrate certain 
 main features,  in particular 
  the role of nonperturbative, gauge-invariantly defined 
 factors associated with infrared subgraphs (both collinear and soft), 
and the role of infrared subtractive techniques 
that serve to identify these factors.   See also~\cite{rogers} for recent 
analyses  along these lines  for  more general processes involving 
fully  unintegrated pdfs. 

One of the questions that a full factorization statement will  address 
is the treatment of soft gluons exchanged between 
subgraphs in different collinear directions. The underlying 
dynamics is that of  non-abelian 
Coulomb phase, treated a long time ago in~\cite{dyproof}  
for the fully inclusive Drell-Yan case.  
But a systematic treatment for more complex observables, including color 
in both initial and final states, is still  missing, as emphasized recently 
in~\cite{collins0708,vogel0708,bomhmuld} 
for   di-hadron and   di-jet hadroproduction near the back-to-back 
region.\footnote{Note 
 that interestingly in~\cite{manch}, which has 
a different point of view than  TMD, 
 Coulomb/radiative mixing terms are 
found to be 
 responsible for the 
breaking of  angular ordering in the initial-state cascade and 
 the appearance of superleading 
logarithms in di-jet cross sections with a gap in rapidity.}

A further question concerns lightcone divergences~\cite{jcc-lc08} 
and the $x \to 1$ endpoint behavior. 
The singularity structure at $x \to 1$  is different  
 in the TMD case than for 
ordinary (integrated) distributions, giving divergences  
 even in dimensional regularization with 
an infrared cut-off~\cite{fhfeb07}.  
 The singularities can be understood 
in terms of gauge-invariant eikonal-line matrix elements~\cite{fhfeb07},   
and the TMD  behavior can be related 
to cusp anomalous dimensions~\cite{korchangle,chered} 
and lack of complete KLN cancellations~\cite{collsud}. 
 In general this  affects the precise form  of factorization and 
relation with collinear  distributions. 
 
Applications  of u-pdfs  at low energies include  
semi-inclusive leptoproduction (\cite{anselm08,ceccopieri,muldetal}, and 
references therein), spin asymmetries~\cite{koike},  
 exclusive reactions~\cite{bochumgpd}. 
In these  cases  
infrared subtractive techniques of the type~\cite{jccfh00,jccfh01} serve for 
   TMD-factorization calculations~\cite{jiyuan} and in particular for 
the proper 
treatment of overlapping momentum regions.\footnote{Subtraction 
techniques related to those  of~\cite{jccfh00,jccfh01}  are 
 developed in~\cite{manohstew} for soft-collinear effective theory, 
and studied in~\cite{leesterm} and~\cite{idimeh} 
in relation with standard perturbative 
methods.  See also SCET applications to  shower algorithms~\cite{bauergen},  
 TMD pdfs~\cite{chay} and jet event shapes~\cite{trottetal} for use of these 
techniques.}   
At high-energy colliders,  
  general characterizations of TMD distributions 
will be  relevant    for turning  
present k$_\perp$-showering generators into general-purpose 
tools to describe hadronic final states over the whole 
phase space~\cite{hj_rec,fhdistalk}.

In the rest of this section   we  will   consider 
applications of k$_\perp$-shower generators to multi-jet final 
states.    
 The main focus is  on regions where jets  are far from  back-to-back, 
 and the total  
energy is much larger than the 
transferred momenta  so that the values of $x$ are small. In this regime 
 the ambiguities   related to 
 soft   Coulomb exchange and to  lightcone divergences  are not 
expected to be crucial.   
We will find that the TMD distributions,  as well  as the 
transverse-momentum dependence of short-distance matrix elements, play 
a very essential role to describe correlations in angle and momentum of 
the  jets.

\subsection{\boldmath{\kt}  shower    with u-pdfs}
\label{sec_implem}

Monte-Carlo event generators based on  
unintegrated pdfs  use   
 factorization at fixed k$_\perp$~\cite{hef} in order to 
a)~generate the 
hard scattering event, including 
  dependence on the initial 
transverse momentum, and 
b)~couple this to the  evolution of the  
initial state to simulate the   parton cascade. 
Implementations of this kind include~\cite{mw92,junghgs,lonn,golec-mc,krauss-bfkl}.  
The hard scattering event is generated by 
\kt-dependent matrix elements (ME)   computed from 
perturbation theory. 
Different generators differ by the detailed model for initial state. 
For the calculations that follow  we use the  Monte Carlo 
implementation \cascade~\cite{junghgs}. 

 The hard  ME 
in the Monte Carlo  are obtained by  perturbative 
calculation~\cite{hef}, while the 
 u-pdfs   are  determined from fits 
to experimental data~\cite{hj04}.  The parton-branching equation used 
  for 
the unintegrated gluon distribution  $ {\cal A} $ 
is schematically  of the form~\cite{mw92,junghgs,hj04} 
\begin{eqnarray}
\label{uglurepr1}
  {\cal A} ( x , \kt , \mu ) & = & 
  {\cal A}_0 ( x , \kt , \mu ) + 
\int { {dz} \over z} \int { { d q^2} \over q^2} \ 
\Theta   (\mu - z  q) \
\nonumber\\
& \times & 
 \Delta    (\mu , z  q) 
\ {\cal P} ( z, q, \kt)   
\   {\cal A} 
 ( { x \over z} , \kt  + (1-z) q, q )
 \hspace*{0.3 cm} .        
\end{eqnarray} 
The first term in the right hand side of Eq.~(\ref{uglurepr1}) 
is the contribution of the 
non-resolvable branchings between  starting scale 
$Q_0$ and  evolution scale $\mu$, 
 and is given by 
\begin{equation}
\label{uglurepr2}
  {\cal A}_0 ( x , \kt , \mu ) =  {\cal A}_0 ( x , \kt , Q_0 ) 
 \ \Delta (\mu , Q_0) 
 \hspace*{0.3 cm}    ,   
\end{equation} 
where $\Delta$ is the Sudakov form factor, and the 
 starting  distribution   
 ${\cal A}_0 ( x , \kt , Q_0 )$     
at scale $Q_0$ is determined from   data fits. 
Details on  the  starting  distribution  used 
for the calculations that follow are  
given   in Appendix~\ref{app:ktfits}. 

The integral term in the right hand side of Eq.~(\ref{uglurepr1}) 
gives the \kt-dependent branchings in terms of the 
 Sudakov form factor $\Delta$ and unintegrated 
  splitting function ${\cal P}$. 
\begin{figure}[htb]
\vspace{37mm}
\includegraphics{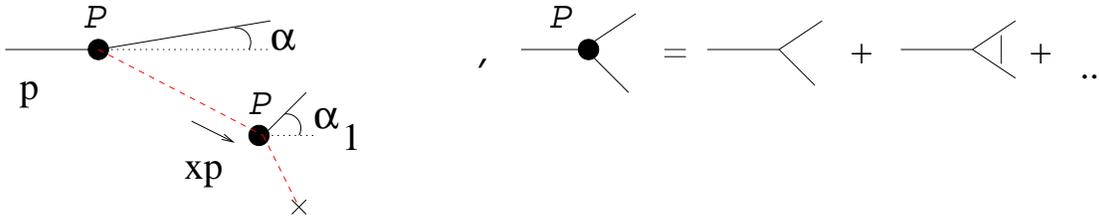}
\caption{(left) Coherent radiation 
 in the space-like parton shower for $x \ll 1$; (right) the unintegrated 
splitting function ${\cal P}$, including small-$x$ virtual 
corrections.} 
\label{fig:coh}
\end{figure}
The explicit expressions  
for these factors  are specified in~\cite{hj04}, and 
include  the effects of coherent gluon radiation not only 
at large $x$ (as e.g. in \herwig) but also 
at  small $x$~\cite{skewang} in the angular region (Fig.~\ref{fig:coh}) 
\begin{equation}
\label{cohregion}
\alpha / x  > \alpha_1 > \alpha \hspace*{0.3 cm}    , 
\end{equation} 
where the angles  $\alpha$ for the 
partons radiated from the initial-state 
shower are taken with respect to the 
initial beam jet direction, and increase with increasing off-shellness. 
Unlike conventional showers, the splitting function  ${\cal P}$ depends on 
transverse momenta and includes part of the virtual corrections, in   such a  way  as 
  to avoid double  counting with the Sudakov form factor while 
 reconstructing  color coherence  in the  small-x  region (\ref{cohregion}).

The Monte Carlo also contains time-like parton showering. 
The impact  of time-like showers  on the jet observables that we will examine in this 
section   turns out to be  very small. This is  similar to what is found in  the 
studies~\cite{d02005,albrow} of 
   Tevatron di-jets and  $\Delta \phi$  distribution, based 
   on \herwig\ and \pythia.   See  remark 
    toward  the beginning of  Sec.~\ref{sec2}. 
Some details on the   treatment  of time-like showering effects 
 are reported  in  Appendix~\ref{sec_timelike}.  
A more complete account  of this topic 
  may be found in~\cite{deaketal}.

\subsection{Azimuthal jet distributions}
\label{sec_numeric}

The \kt-dependent ME and parton branching 
lead to a different angular pattern of 
initial-state gluon radiation   
than standard, collinear-based showers, e.g. \herwig. 
In particular, while the \herwig\  
angular ordering  reduces  to  ordering  
in transverse momenta for x~$\to$~0, 
 the \kt-dependent shower contains finite-angle 
corrections in this limit~\cite{boand02}. 
 We now compute angular 
distributions for the ep  three-jet cross section 
by   the  {\kt}-shower   Monte Carlo \cascade\ and    by \herwig.   
\begin{figure}[htb]
\vspace{140mm}
%\includegraphics[width=8.2cm]{zeus-delta-phi-3jet-low.eps}
%\includegraphics[width=8.2cm]{zeus-delta-phi-3jet-hig.eps}
%\vspace*{-5.8cm}
\includegraphics{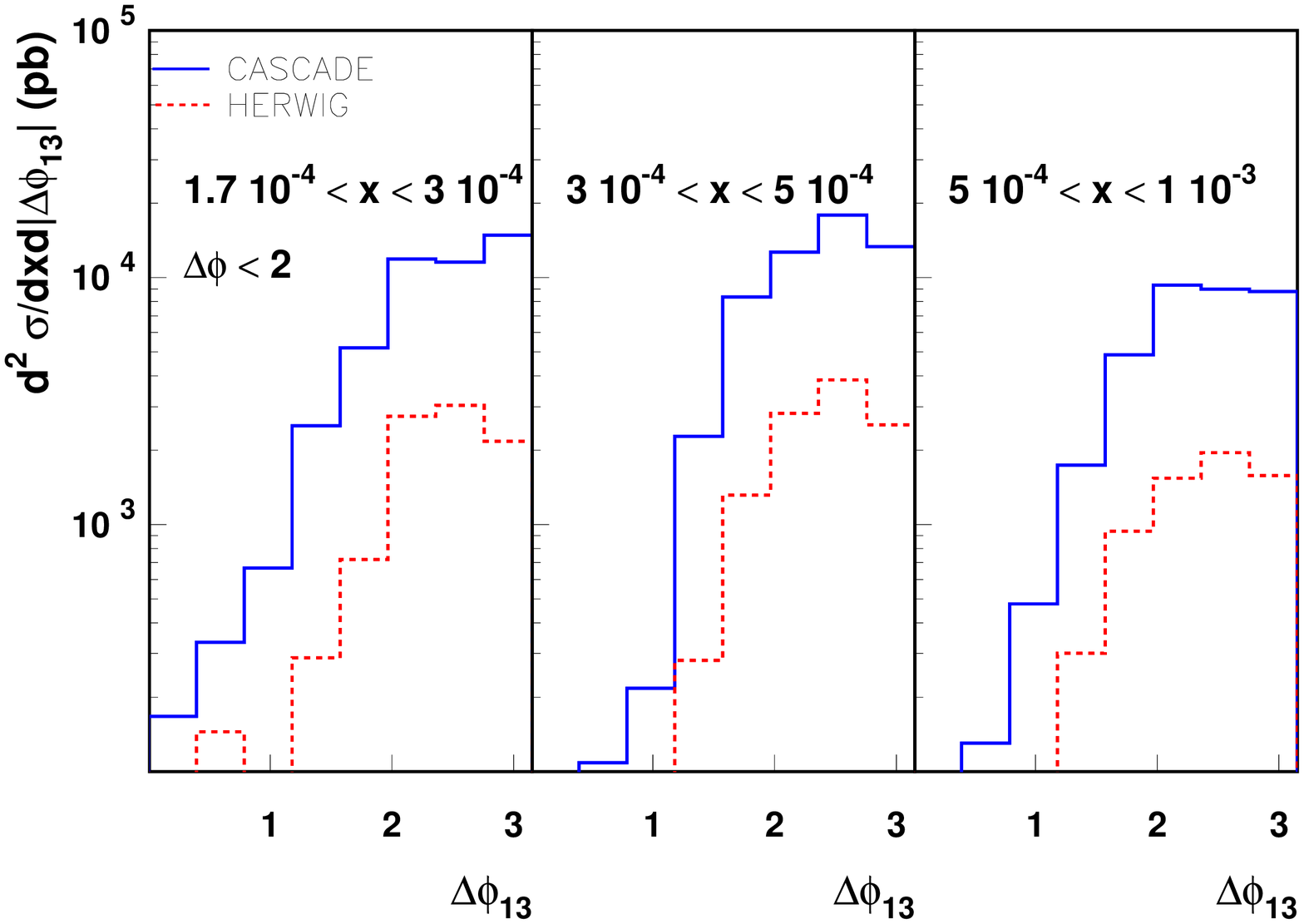}
\includegraphics{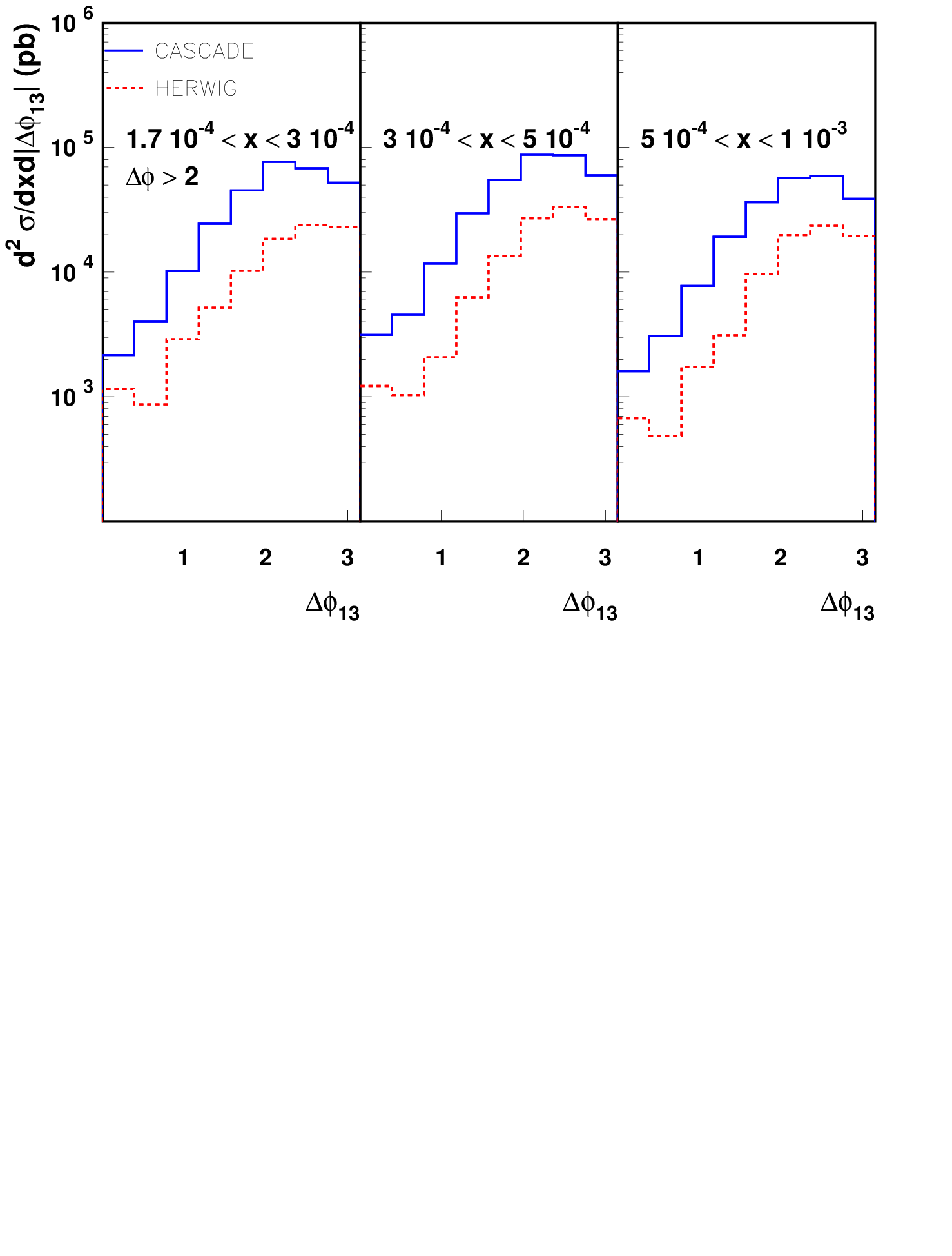}
\caption{Cross section in the azimuthal angle 
$\Delta \phi_{13}$ between the 
 hardest and the
3rd jet for  small
($\Delta \phi < 2$, top) and 
large ($\Delta \phi > 2$, bottom)  azimuthal  
separations between the leading jets. The \kt  
Monte Carlo results \cascade\ are compared with \herwig.
} 
\label{fig:thirdjet}
\end{figure}
Let $\Delta \phi$ be    the azimuthal separation 
between the two jets with the highest transverse energy $E_T$, 
\begin{equation}
\label{deltaphi}
\Delta \phi = \phi_{\rm{jet-1}} - \phi_{\rm{jet-2}}  \hspace*{0.1 cm} ,    
\end{equation}
where the azimuthal angle $\phi$ 
for each jet is defined in the 
hadronic center-of-mass frame. 
%as the $E_T$-weighted average of the 
%angles $\phi_n$ for the  momenta $\{ p_n \}$ clustered 
%in the jet, 
%\begin{equation}
%\label{jetphi}
% \phi = {{ \sum_n \phi_n \ E_{T n}}  \over  {\sum_n  E_{T n}} }   
% \hspace*{0.1 cm} . 
%\end{equation}
Similarly, we define $\Delta \phi_{13}$ as 
the azimuthal separation between the 
 hardest and the
third jet. 

In Fig.~\ref{fig:thirdjet} we compute  
the three-jet 
cross section and plot it versus  the azimuthal angle 
$\Delta \phi_{13}$, by distinguishing the cases 
in which the two leading jets are at small 
angular separation ($\Delta \phi < 2$) or large 
angular separation ($\Delta \phi > 2$). 
 \cascade\  gives large differences from  \herwig\
in the region where the azimuthal separations  
 $\Delta \phi$ between the 
leading jets are small, see top plot of 
Fig.~\ref{fig:thirdjet}. 
\begin{figure}[htb]
\vspace{143mm}
%\includegraphics[width=8.2cm]{zeus-delta-phi-dijet.eps}
%\includegraphics[width=8.2cm]{zeus-delta-phi-3jet.eps}
%\vspace*{-6.9cm}
\includegraphics{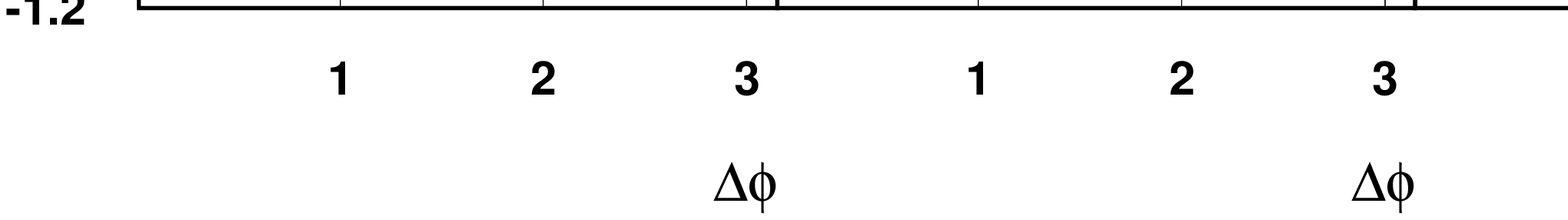}
\includegraphics{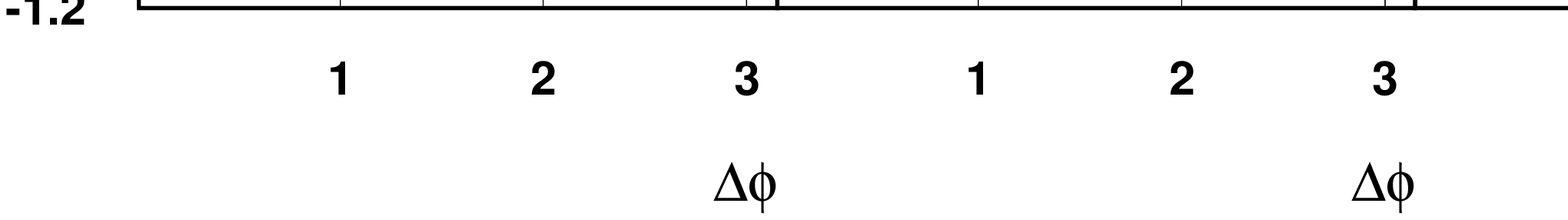}
\caption{Angular jet correlations  obtained by the \kt-shower 
\cascade\ and by \herwig, compared 
with $e p$  data~\protect\cite{zeus1931}: 
(top) di-jet cross section; (bottom) three-jet cross section. 
The \herwig\    results are multiplied by a factor of 2. 
} 
\label{fig:phipage}
\end{figure}
This reflects the fact that at small 
 $\Delta \phi$ the phase space opens up for events in 
which the partonic lines along the initial decay chain are not 
ordered in transverse momentum.  Such configurations are 
taken into account in \cascade\ with the appropriate 
matrix element, at least for small enough $x$,   but not in \herwig.  
   The $x$ values considered 
in  Fig.~\ref{fig:thirdjet} are those corresponding to the 
three-jet measurements in~\cite{zeus1931}. 
As  $\Delta \phi$ increases, the results from 
\cascade\ and  \herwig\ 
become closer. See  bottom plot of 
Fig.~\ref{fig:thirdjet}.  This is  associated with the fact that 
for $\Delta \phi$ approaching the 
back-to-back region  the phase space for finite-\kt emissions 
is reduced. In this region one thus expects 
 both Monte Carlos  to give  reasonable approximations.

 Fig.~\ref{fig:phipage} shows the angular correlations for 
    final states with two jets and three jets. 
We compute the azimuthal distribution  of di-jet and three-jet 
cross sections 
in  the separation  $\Delta \phi$ between the leading jets. 
We show the   distributions  obtained by 
\cascade\ and by \herwig,   compared with the 
measurement~\cite{zeus1931}. 
We multiply the \herwig\   result by a constant factor equal to 2, which the top plot 
in  Fig.~\ref{fig:phipage} shows is the K-factor needed in order to get 
the normalization approximately correct for the two-jet  region. 
Observe that the shape of the distribution is different for the two 
Monte Carlos. As expected from the result of Fig.~\ref{fig:thirdjet}, 
\cascade\ gives the largest differences to  \herwig\ 
at  small $\Delta \phi$, and becomes closer to \herwig\  as 
$\Delta \phi$ increases. 
In particular, we observe that while the K-factor of 2 for  \herwig\  is sufficient for the 
two-jet region, the  shape of the jet  distribution is not  properly 
described  by    \herwig\   as  $\Delta \phi$ decreases. 
The description of the 
measurement by \cascade\ is
 good, whereas  \herwig\ is not
sufficient to describe the measurement 
in the small $\Delta \phi$ region.
 We  further  see  in the bottom plot of   Fig.~\ref{fig:phipage}   
that the three-jet cross section is reasonably well described by 
the    {\kt}-shower  result but not   by   \herwig.

\begin{figure}[htb]
\vspace{90mm}
\includegraphics{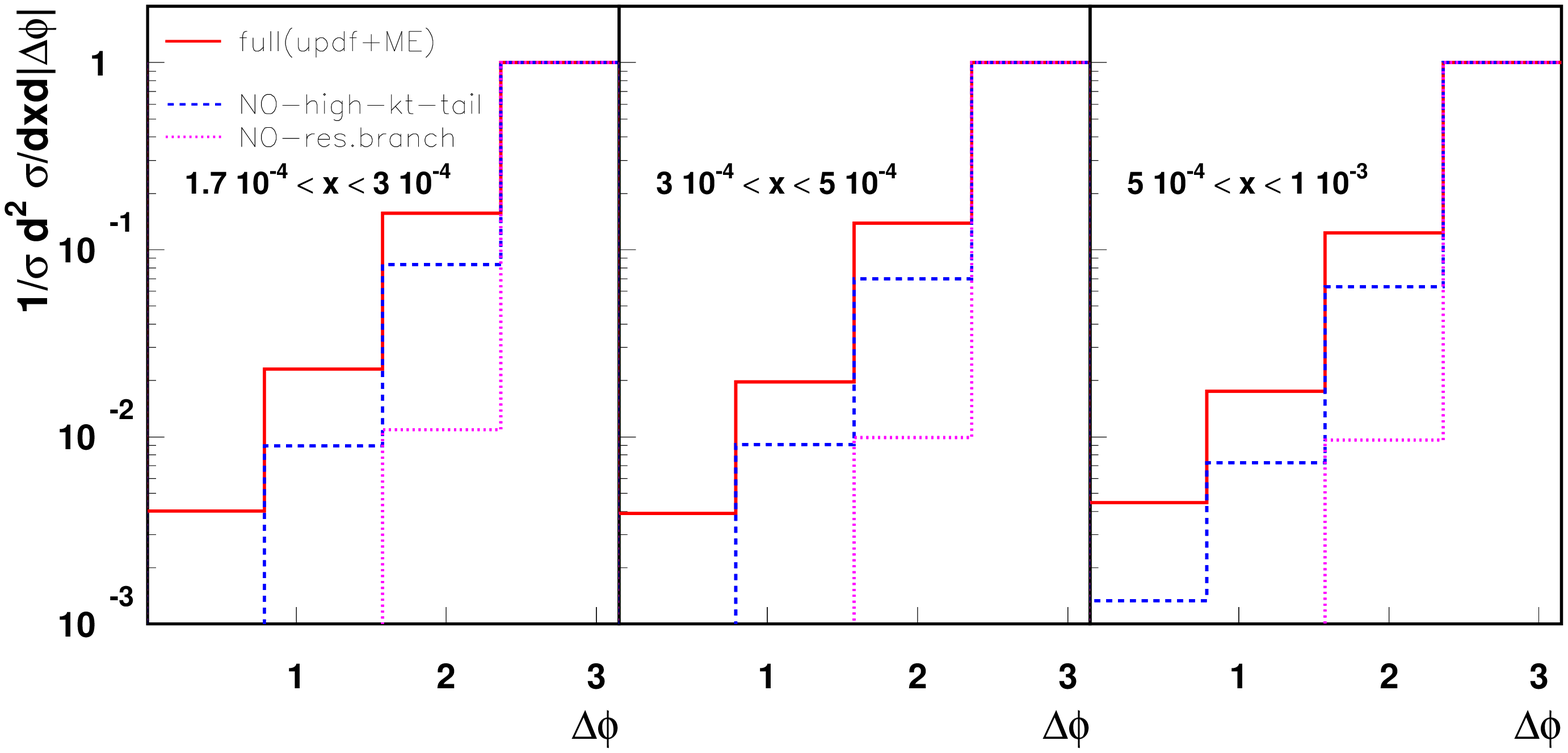}
\caption{Azimuthal distribution normalized to the
back-to-back cross section:
(solid red)  full result
(u-pdf $\oplus$ ME); (dashed blue) no finite-\kt
correction in ME
 (u-pdf $\oplus$ ME$_{collin.}$);
(dotted violet) u-pdf with no resolved branching.
}
\label{fig:ktord}
\end{figure}

Note that the interpretation  of
the angular correlation data
in terms of corrections to collinear ordering  is
 consistent  with the finding~\cite{zeus1931} discussed 
 in Sec.~\ref{sec2} that 
while inclusive jet rates are reliably predicted by NLO fixed-order
results, NLO predictions
are affected by  large corrections to di-jet azimuthal distributions
(going from ${\cal O} (\alpha_s^2)$ to ${\cal O} (\alpha_s^3)$) in the
small-$\Delta \phi$ and small-$x$ region, and begin to fall below the
data for three-jet distributions  in the 
smallest $\Delta \phi$ bins (Fig.~\ref{fig:phizeus-3j}~\cite{zeus1931}).

The  physical picture underlying    the \kt-shower calculation
 in Figs.~\ref{fig:thirdjet},\ref{fig:phipage} involves  both  
transverse-momentum dependent  parton distributions (determined from
  experiment) and  
  matrix elements (computed perturbatively).
Fig.~\ref{fig:ktord} illustrates  the relative contribution of these   
different components to   the   result, 
showing different approximations to the
azimuthal dijet distribution normalized to the
back-to-back cross section. The solid
red curve is   the full  result. 
  The dashed blue curve is  obtained
from the same unintegrated pdf's but
by taking the collinear approximation in
the hard matrix element, 
\begin{equation}
\label{mcoll}
{\cal M} ( \kt ) \to {\cal M}_{collin.} ( \kt )  = 
{\cal M} ( 0_\perp ) \ \Theta ( \mu - \kt ) 
 \hspace*{0.1 cm} . 
\end{equation} 
The dashed curve  
drops much faster than the full result as $\Delta \phi$ decreases, 
indicating  that  the
high-\kt component  in the hard ME~\cite{hef} 
is necessary  to describe 
jet correlations      
for small $\Delta \phi$~\cite{radcor}.  For reference we also 
plot, with the dotted (violet) curve,  the result 
  obtained from the 
unintegrated pdf 
without any resolved branching, 
\begin{equation}
\label{uglunores}
  {\cal A} ( x , \kt , \mu ) \to 
 {\cal A}_{no-res.}  ( x , \kt , \mu ) = 
  {\cal A}_0 ( x , \kt , Q_0 ) 
 \ \Delta (\mu , Q_0) 
 \hspace*{0.3 cm}    .     
\end{equation} 
Here 
 ${\cal A}_0$ is the starting  distribution at $Q_0$ 
and $\Delta$ is the Sudakov form factor, giving the 
no-radiation probability between $Q_0$ and $\mu$.  
This   represents   the contribution of 
the intrinsic \kt  distribution only, 
corresponding
 to nonperturbative, predominantly
 low-\kt modes.  That is,  in the dotted (violet) curve one retains 
an  intrinsic 
 \kt $\neq 0$ but no  effects of coherence. We see  
 that the resulting jet correlations in this case are down by an order of magnitude.

The results of
Fig.~\ref{fig:ktord}  illustrate  that the  
 \kt-dependence  
in the unintegrated pdf alone
is not sufficient  to describe jet production
quantitatively, and that  
 jet correlations are  sensitive to the 
 finite, high-\kt tail of matrix elements~\cite{hef}  
computed  from perturbation theory. 
We note  that the inclusion of  the  perturbatively computed  high-k$_\perp$
 correction
 distinguishes the  present calculation  of  multi-jet cross sections
 from other  shower approaches (see
  e.g.~\cite{krauss-bfkl})
 that include transverse momentum
dependence in the  pdfs but not  in the  matrix elements.

\vspace*{4.6 cm}
\begin{figure}[htb]
%\special{psfile=jj_fig1.ps angle=-90 hscale=25 vscale=25
%  hoffset=-210 voffset=140}
%\special{psfile=jj_fig2.ps angle=-90 hscale=25 vscale=25
%  hoffset=20 voffset=140}
\includegraphics{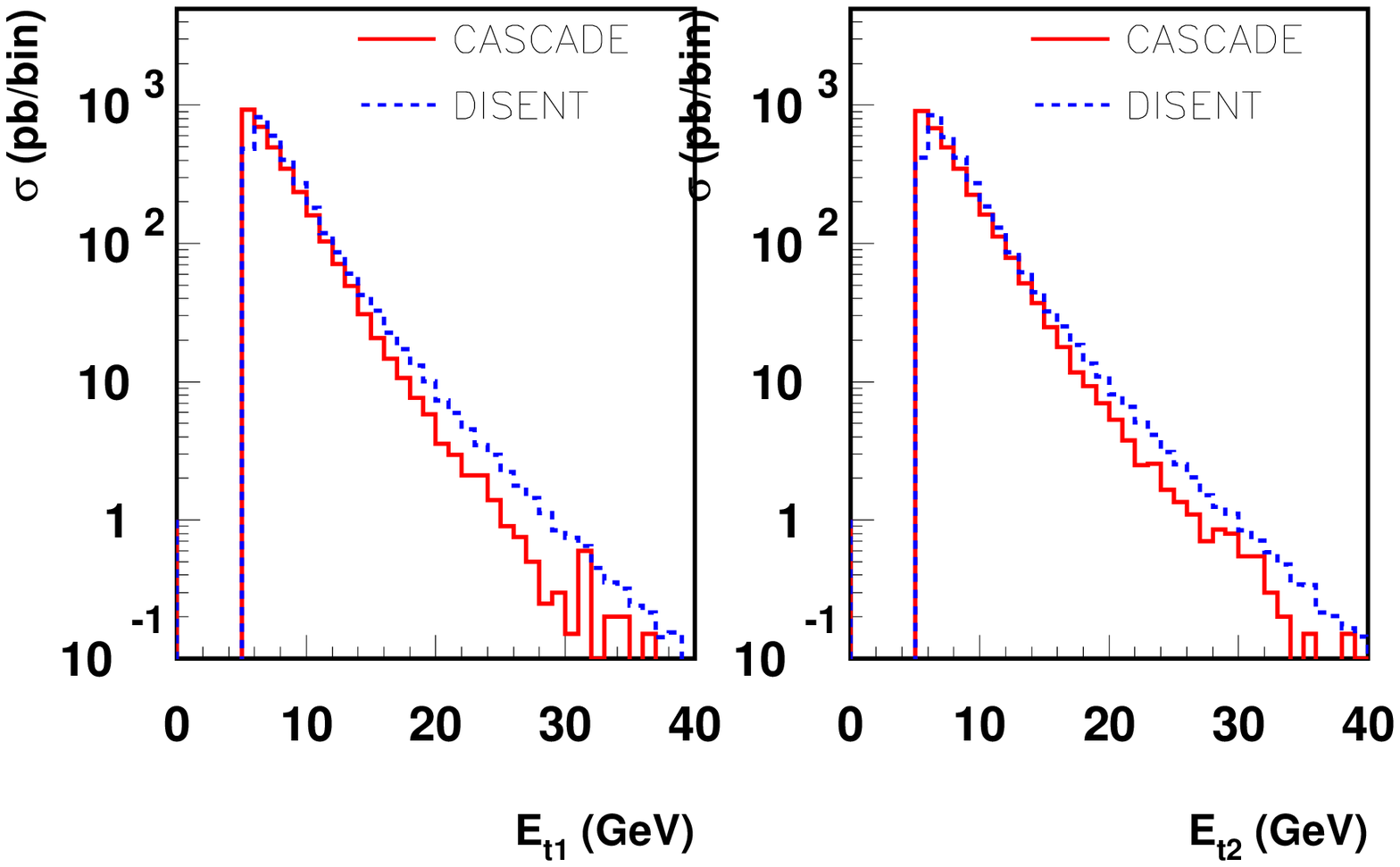}
\includegraphics{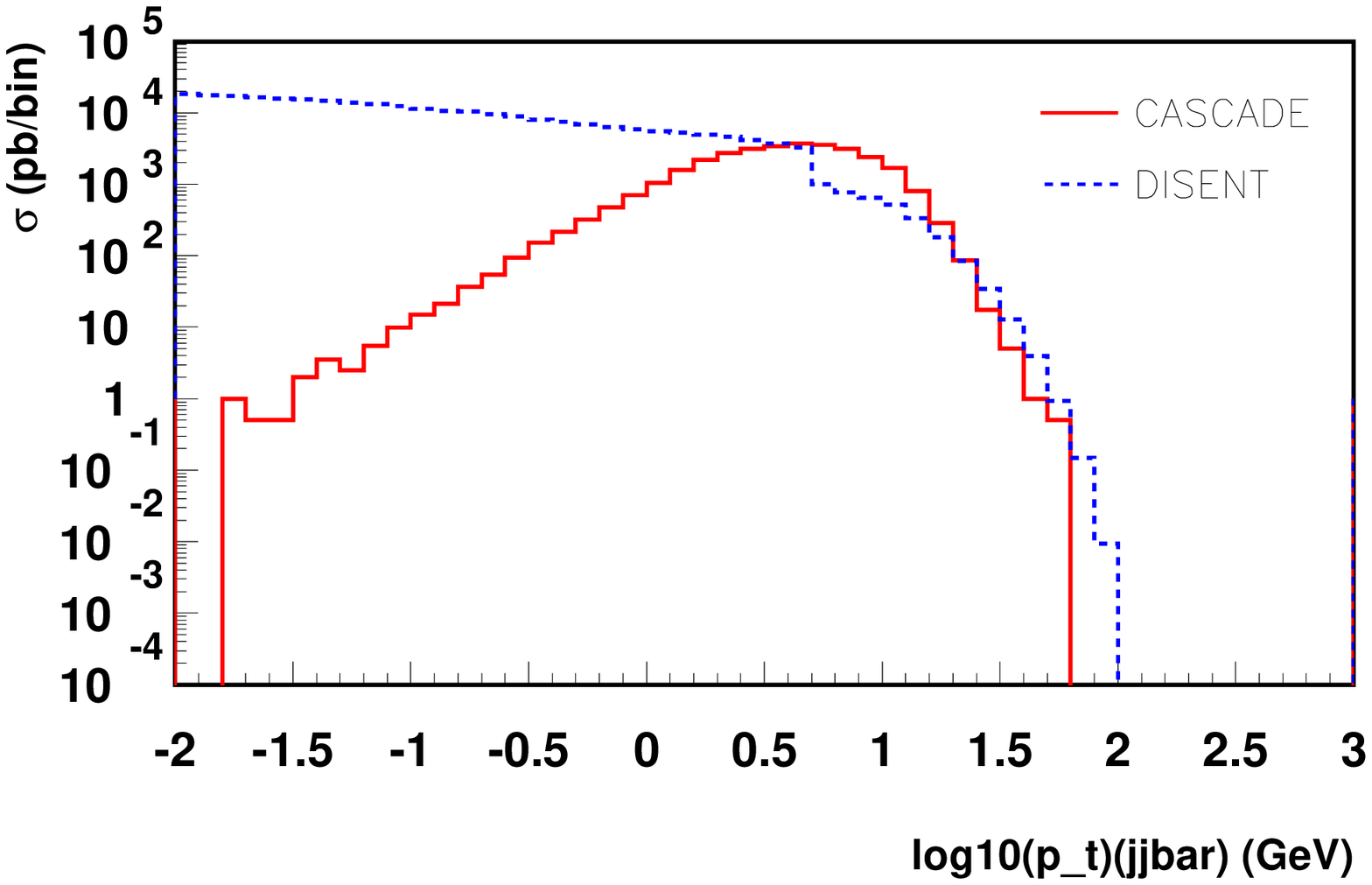}
\caption{Comparison of the k$_\perp$-shower  \cascade\
with the  NLO di-jet calculation \disent: (left)~distribution
in single-jet transverse energy; (right)  distribution in the
di-jet transverse energy.}
\label{fig:jj}
\end{figure}

To examine more closely  the distribution in k$_\perp$ that results
from highly off-shell
subprocesses, in  Fig.~\ref{fig:jj} we study the jet cross section
in transverse energy and compare the  k$_\perp$-shower with the NLO
 result from the \disent\ event 
 generator. It is noteworthy that the  large-$p_t$ part of the di-jet
spectrum is very close for the two calculations. At low $p_t$ one
sees the Sudakov form-factor effect in the shower result.
Differences in the single-jet spectra are also of interest 
and can be shown to be   associated  with  quark contributions~\cite{deaketal}. 
These detailed 
comparisons  may be of use to relate~\cite{dasgqt}
{\small DIS} event shapes measuring
the transverse momentum  in the current region
 to hadro-production $p_T$ spectra.

\begin{figure}[htb]
\vspace{60mm}
%\special{psfile=jj_fig1.ps angle=-90 hscale=25 vscale=25
%  hoffset=-210 voffset=140}
%\special{psfile=jj_fig2.ps angle=-90 hscale=25 vscale=25
%  hoffset=20 voffset=140}
\includegraphics{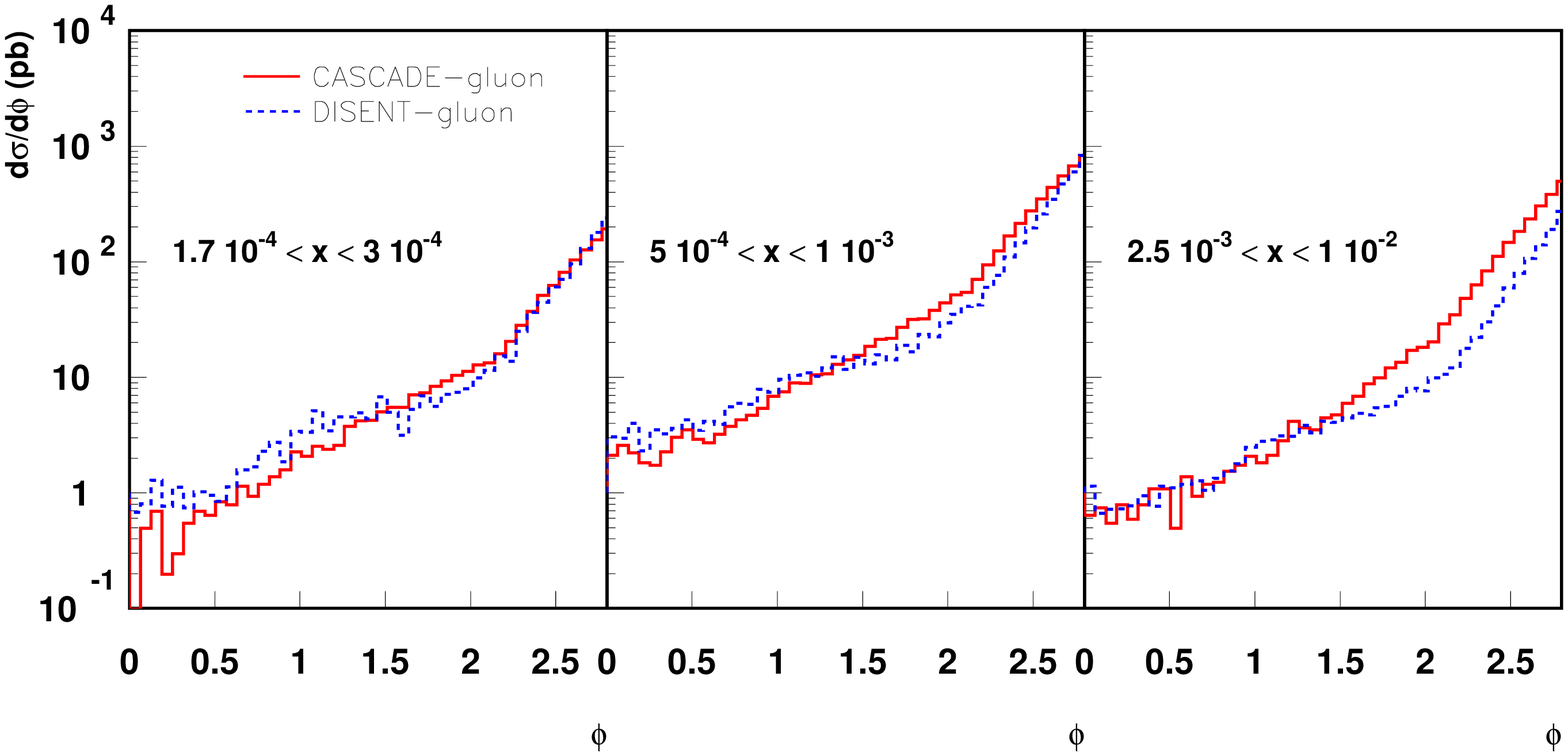}
\caption{Azimuthal di-jet distribution obtained from the 
expansion of the k$_\perp$-shower   \cascade\  to one-gluon emission level, 
compared with the  NLO di-jet calculation \disent  (gluon  channel).}
\label{fig:one-glu}
\end{figure}

In Fig.~\ref{fig:one-glu} we push  
 further   the comparison at next-to-leading-order level. 
 We switch off hadronization, and use the k$_\perp$-shower Monte Carlo  \cascade\   
 as a parton-level  generator.  
 We evaluate the k$_\perp$-shower by expanding in the number of extra emissions, and 
 truncate to the level of one gluon emission. We compare this with the    
NLO    \disent\   calculation, taking  only  the  gluon channel   in     \disent. 
We compute  the azimuthal di-jet distribution at  various  values of  x. 
The plots in Fig.~\ref{fig:one-glu} indicate that for sufficiently small x 
the one-gluon expansion  of  the shower program     agrees with  the full  NLO 
 result.  
We view this as a numerical consistency check of the shower program in the case 
of  a relatively complicated final-state  correlation, to be considered jointly with the 
analytic cross-checks     quoted in 
Sec.~\ref{sec2prime},   e.g.~\cite{vannee},   
 for  the case of   analytic small-x results for inclusive  observables.

\begin{figure}[htb]
\vspace{180mm}
\includegraphics{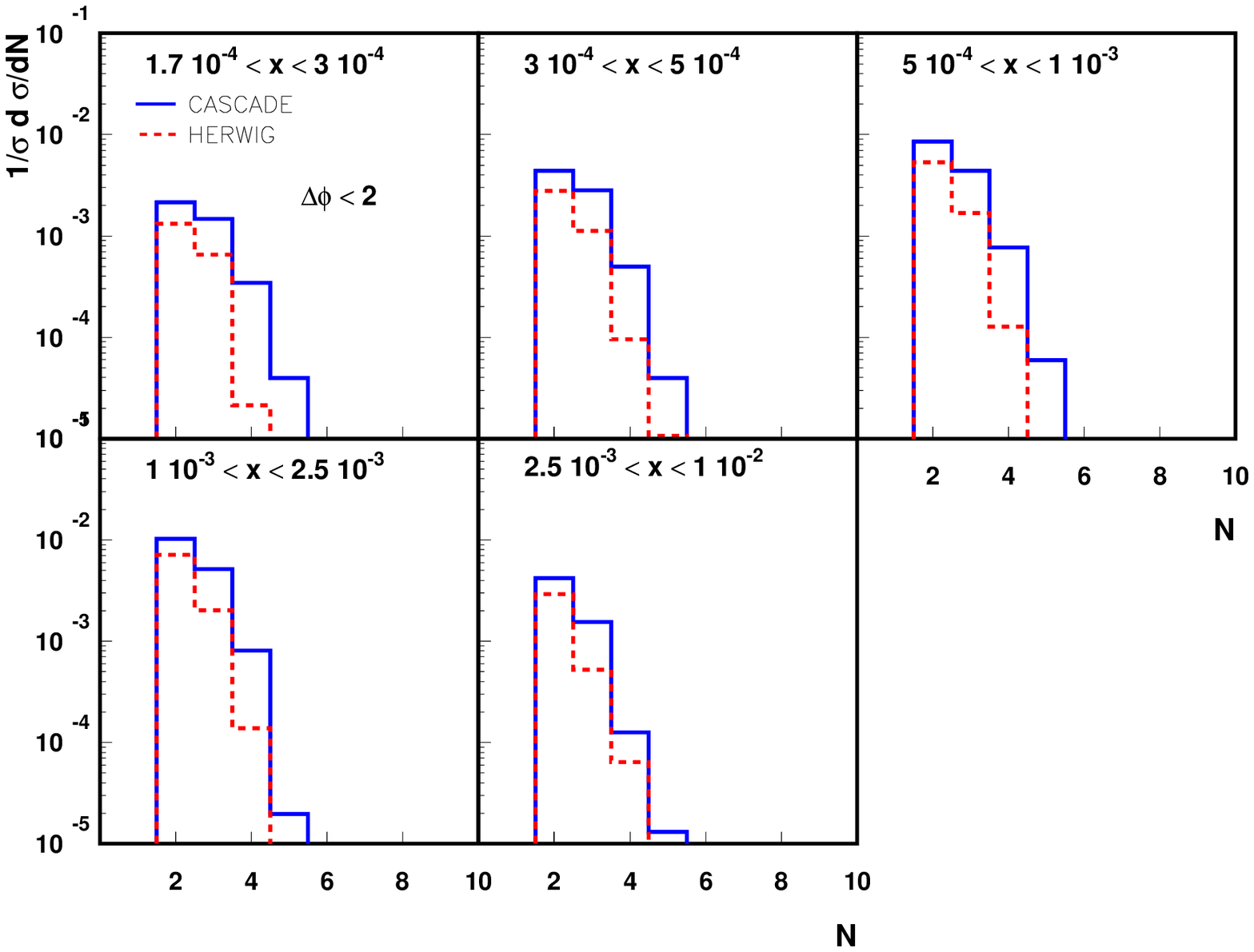}
\includegraphics{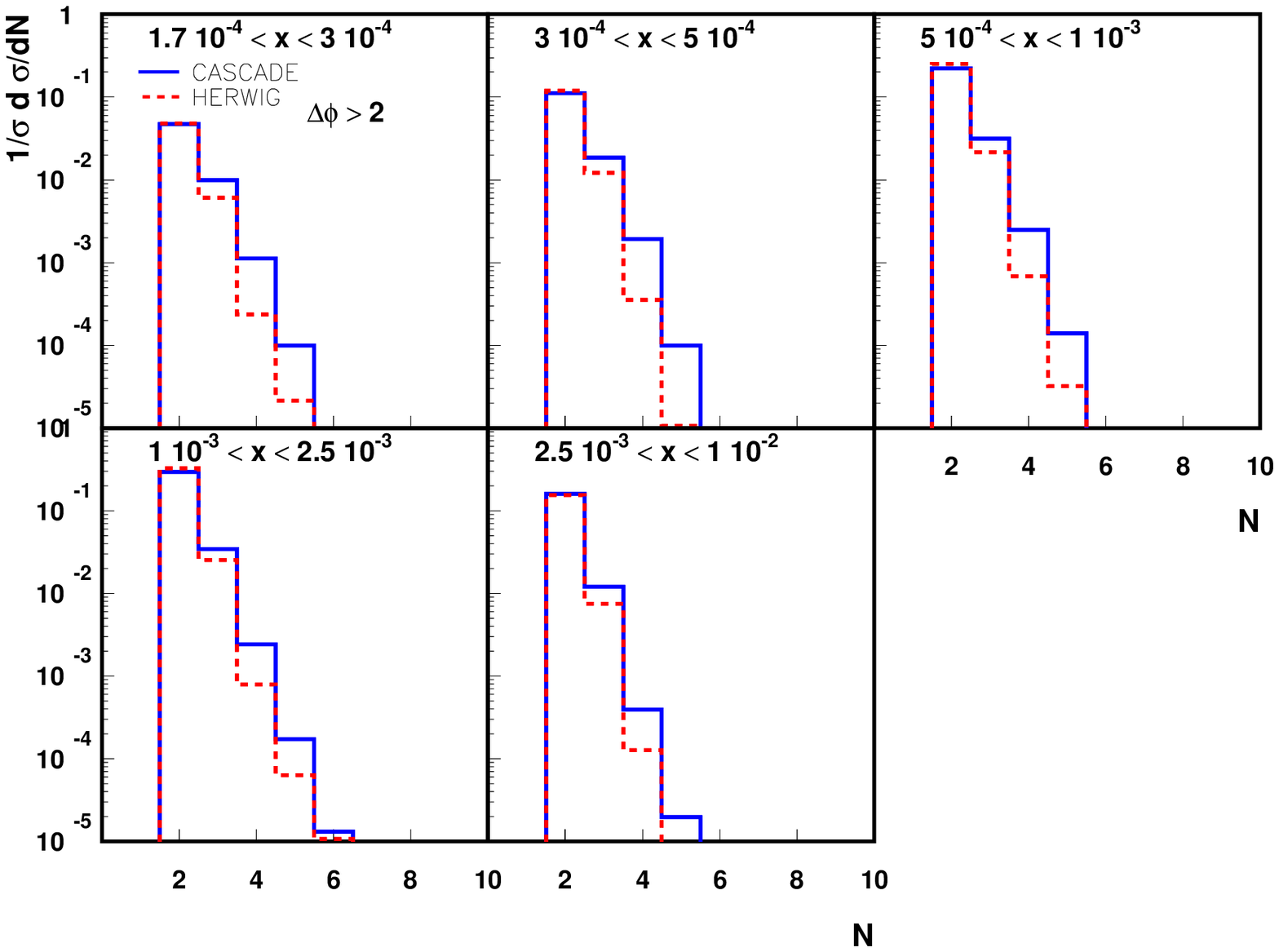}
\caption{Jet multiplicities  obtained by \cascade\ and \herwig \ for  
(top) $\Delta \phi < 2$ and  
(bottom) $\Delta \phi > 2$.
} 
\label{fig:jmult}
\end{figure}

We conclude this section by 
observing  that  the jets  that we are 
considering  are produced 
in the region of rapidities   of Eq.~(\ref{kinetaet}), away from 
  the forward region.   While 
forward-region observables are  relevant  
in their own right and have long been studied as  probes 
of the initial-state shower dynamics 
(see e.g.~\cite{heralhcproc,jeppe04} and references therein), 
Monte Carlo results for such observables have a more pronounced  
 dependence on the details of the model 
used for u-pdf evolution~\cite{boand02} (see also 
discussions  in~\cite{rogers,ceccopieri,collins01}). It is thus interesting that  
significant  effects of non-ordering in \kt \ 
for the space-like shower 
are found in the present case for centrally produced hard jets.

\section{Jet multiplicities and momentum correlations}
\label{sec4}

We now turn to  jet multiplicities and 
 transverse-momentum correlations. These observables  provide 
further details on the 
structure of the multi-jet final states. As noted in 
Sec.~\ref{sec2}, several of the transverse-momentum correlations 
measured  in~\cite{zeus1931} are affected by sizeable theory 
uncertainties at NLO~\cite{zeus1931,nagy}.

\begin{figure}[htb]
\vspace{150mm}
\includegraphics{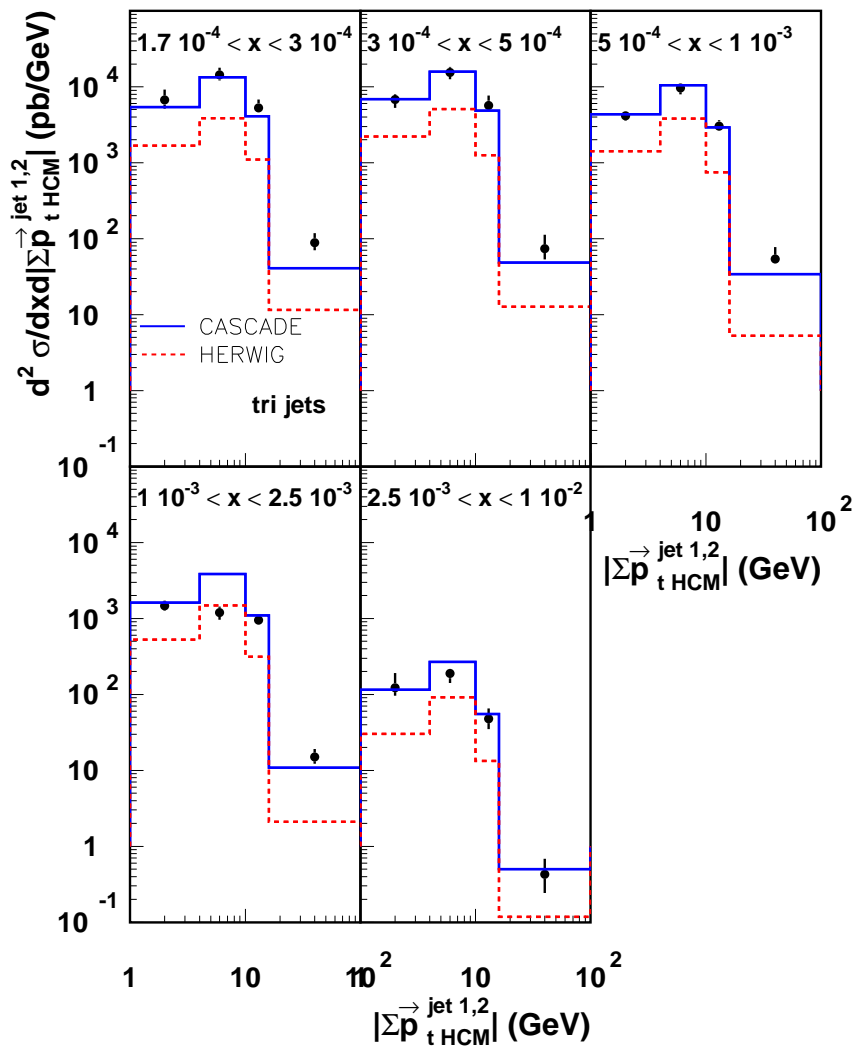}
\caption{Momentum correlations  obtained by 
\cascade\ and \herwig, compared 
with $e p$  data~\protect\cite{zeus1931}: 
 three-jet cross section versus 
%$| \sum p_T^{1,2} |$, 
% the magnitude of the sum of the $p_T$ for the two highest-$E_T$ jets.  
the variable $| \sum p_T^{1,2} |$ introduced in the text.    
} 
\label{fig:momcorr1}
\end{figure}

Let us  first consider  jet multiplicity distributions.   
Finite-\kt \   corrections  increase the 
mean gluon multiplicity and broaden the 
spectrum~\cite{skewang,hef,mw92}. 
In Fig.~\ref{fig:jmult} we compute  
 the distribution in the number of jets $N$, normalized to the 
two-jet cross section $\sigma$. 
As in Fig.~\ref{fig:thirdjet}  
we show  
separately  the results for small and large azimuthal 
separations between the hardest jets, $\Delta \phi < 2$ and 
$\Delta \phi > 2$. 
Jet multiplicities at small $\Delta \phi$ are where the 
clearest differences appear between the two parton showers. 
The \kt-shower result  receives  larger 
contribution from high multiplicities. 
Besides the absolute size of this contribution,   note that 
 Fig.~\ref{fig:jmult} illustrates  the 
difference in the  shape  between the 
two Monte Carlos. 

\begin{figure}[htb]
\vspace{150mm}
\includegraphics{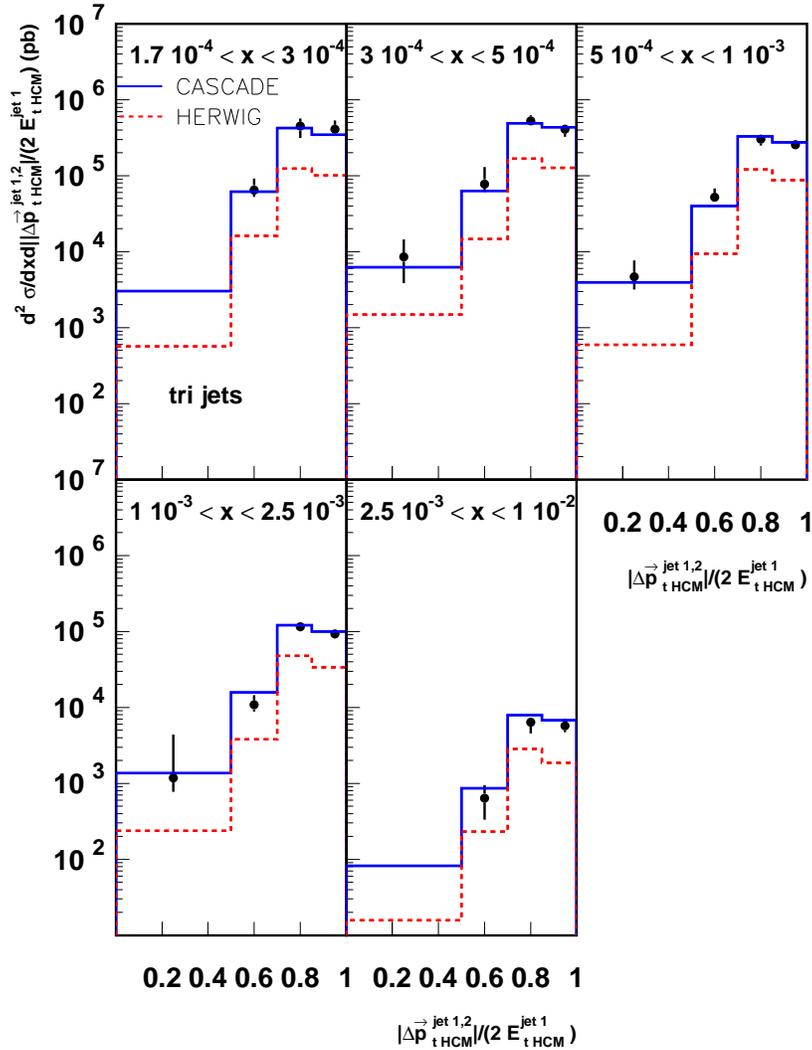}
\caption{Momentum correlations  obtained by 
\cascade\ and \herwig, compared 
with $e p$ data~\protect\cite{zeus1931}: 
  three-jet cross section versus 
%the vector difference of the highest-$E_T$ jet transverse momenta, 
%scaled by twice the transverse energy of the hardest jet. 
the variable 
$| \Delta p_T^{1,2} | / (2 E_T^1)$ introduced in the text.  
} 
\label{fig:momcorr2}
\end{figure}

 In Ref.~\cite{zeus1931}  the {\small ZEUS} collaboration
has  presented measurements of various momentum correlations. 
We examine two such  distributions for three-jet 
cross sections in Figs.~\ref{fig:momcorr1} and \ref{fig:momcorr2}. 
In Fig.~\ref{fig:momcorr1} is shown the distribution in the 
magnitude of the sum of the transverse momenta $p_T$ for 
 the two jets with the highest 
$E_T$,  $| \sum p_T^{1,2} |$. 
The back-to-back region corresponds to $| \sum p_T^{1,2} | \to 0$ in this 
plot. The region of large  $| \sum p_T^{1,2} |$   is the 
region with at least three well-separated hard jets.   The  
\kt-shower  result  
describes this region reasonably  well. The 
results  from  \herwig\  are  quite lower.

In Fig.~\ref{fig:momcorr2} is 
 the distribution  in the vector difference 
of the highest-$E_T$ jet transverse momenta, 
scaled by twice the transverse energy of the hardest jet, 
$| \Delta p_T^{1,2} | / (2 E_T^1)$. 
The back-to-back region corresponds to 
$| \Delta p_T^{1,2} | / (2 E_T^1) \to 1$ in this plot. 
The behavior  of the Monte Carlo results compared to  the 
data  is rather similar to that in 
 Fig.~\ref{fig:momcorr1}.

In summary, the calculations of this paper show that 
the \kt-shower
results describe well the shape of  multi-jet distributions observed
 experimentally, including correlations, and give  quite
  distinctive features of the associated  distributions
  compared  to  standard showers such as  \herwig.
The largest differences between the two parton showers occur when the
azimuthal separations between the leading jets are small, whereas the
results become more similar in  the two-jet region. 
See e.g. Figs.~\ref{fig:phipage},\ref{fig:jmult}. 
In the region of small azimuthal distances
the largest variation occurs   between
order-$\alpha_s^2$ and order-$\alpha_s^3$ results in the
fixed-order NLO calculations, particularly
for small $x$. In cases where corrections
 are not  large,
the  NLO and \kt-shower
 calculations are  rather close. 
The results support a physical
picture of  multi-jet correlations
in which
sizeable radiative corrections arise not only
from collinear/soft emission,  included
   in  \herwig\ as well, but also from
 finite-angle emission, associated
with  the growth of
transverse momenta transmitted along  the space-like jet.
 Small-x coherence effects, computed in this section 
 and the previous section   for jet multiplicities, momentum correlations and 
  angular correlations,    are  included in the  \kt-shower but not in \herwig.  
They  are associated with  multi-gluon radiation terms to the  initial-state shower  
that become
non-negligible at
  high energy   and small 
  $\Delta \phi$.\footnote{Near 
the  back-to-back  region  of large $\Delta \phi$, on the other hand, 
 corrections due to  mixed Coulomb/radiative terms
  can  also  become  important~\cite{collins0708,vogel0708}.}

The above observations suggest the usefulness of combining
NLO and \kt-shower
for a broad range of multi-jet observables, in order to obtain more
 reliable predictions over a wider kinematic region.
Monte Carlo results depend on
  the  maximum angle
parameter $\mu$~\cite{mw92,boand02,junghgs} at which  
the shower is  evaluated.
The perturbative matching will involve this angle.
Studies of the dependence  of  Monte Carlo results
 on  $\mu$ will be reported elsewhere.

\section{Prospects for LHC final states and conclusions}
\label{sec:concl}

Experimental analyses of multi-particle final states at   the Large Hadron Collider   
depend  on realistic   parton-shower Monte Carlo simulations. 
Multi-particle  production  acquires  qualitatively new features at the 
LHC   compared to previous  hadron-hadron  experiments 
due to the large phase space opening up for events characterized by multiple hard 
scales,  possibly widely disparate from each other. 
This brings in both potentially large radiative corrections and potentially new effects 
in the nonperturbative components of  the  process being probed near   
phase-space  boundaries.   
It is not at all obvious that the approximations involved in standard Monte Carlo 
generators that have successfully served for event simulation in past  collider 
experiments will be up to the new situation. 

In this paper 
we have discussed the method of  k$_\perp$-dependent Monte Carlo shower, based 
on transverse-momentum dependent (TMD), or unintegrated, parton 
distributions and matrix elements defined by high-energy factorization. 
 The main advantage of the method over standard Monte Carlo generators 
 is the inclusion of  corrections  to collinear-ordered showers,  
 and of effects of  QCD coherence 
  associated with finite-angle radiation from space-like partons carrying 
  arbitrarily soft    longitudinal momenta.    
  Sensitivity to these dynamical features is bound to be enhanced by the 
  high-energy  multi-scale kinematics.    
 The theoretical basis of the k$_\perp$-shower   method allows one to 
go to arbitrarily high transferred-momentum scales, 
thus making it suitable for   event  simulation of   jet physics  at the LHC. 

In the paper  we have pointed to  developments of the approach 
toward general-purpose event generators, and  illustrated  validation of 
  k$_\perp$-shower   Monte Carlo 
  using   experimental    ep data for final states with multiple hadronic jets.  
  We have noted that while Tevatron di-jet correlations are dominated  
  by leading-order processes, and are  reasonably well described by 
  collinear-based event  generators, this is not so in the   case of ep data.  
  We have found that 
including
finite-k$_\perp$ radiative contributions 
 in the initial state shower   
gives sizeable effects and improves significantly 
the description of 
 angular  correlations and transverse-momentum  
 correlations.  
   Despite the  lower 
ep energy,   the  multi-jet  kinematic region considered   is 
 characterized by   the   large phase space available for jet production   
 and relatively 
small values of the ratio 
between the jet transverse momenta and center-of-mass energy, and 
   is    thus     relevant 
 for     extrapolation    of  initial-state showering effects   to the LHC. 

Besides jet final states, 
the   corrections to collinear-ordered  showers   that we are 
treating  will 
 also affect  heavy mass production at the LHC, 
including  final states with heavy bosons and heavy flavor. 
An example is provided by bottom-quark  pair production. 
Going from the Tevatron to the LHC~\cite{baines}   implies 
a sharp  increase in the relative fraction of events 
dominated by the $g \to b {\bar b}$   subprocess 
coupling~\cite{hef}  to the spacelike jet. 
This  is  bound to affect  the   reliability   of   
shower calculations based on collinear ordering  (as well as the  stability 
 of  NLO perturbative predictions), as these 
 do not properly account for contributions 
of  $ b {\bar b}$  in association with two   hard  jets, with $p_t$ of the heavy quark pair 
large  compared 
 to the bottom-quark  mass but small  compared 
   to the transverse momenta of  the individual associated jets.  
These kinematic  regions are the   analogue of the  regions unordered in 
k$_\perp$   studied in this paper for jet correlations.  
The  fraction  of   $ b {\bar b}$ events of this kind is  
not  very significant   at the Tevatron but will be 
sizeable   at the LHC. 
  The quantitative importance of 
   unordered configurations coupling to $g \to b {\bar b}$ 
  will reduce  the  numerical  stability of collinear-based predictions 
  (NLO,     or parton-shower, or their combination~\cite{mcatnlo})  
  with respect to renormalization/factorization scale  variation in the case of LHC. 
  On the other hand,   these are   precisely 
  the configurations that the k$_\perp$  Monte Carlo shower  is designed 
  to  treat.

Even more complex multi-scale effects are to be expected,  and are 
  beginning  to be investigated~\cite{deaketal},    in 
the associated production of  bottom quark pairs and W/Z bosons~\cite{mlm93}, 
and possibly in final states with Higgs bosons~\cite{hj04,higgs02}\footnote{Non-negligible 
numerical effects of high-energy subleading terms 
were noted~\cite{vogelhiggs} in the predictions for the 
 Higgs transverse-momentum spectrum at the LHC.} especially for 
 measurements of   the  
 less inclusive distributions and correlations.    
 The vector boson case is  relevant for 
  early phenomenology at the LHC, as 
  small-$x$ broadening of W and Z $p_T$ 
 distributions~\cite{olness} (see~\cite{cpyuan1}) affects the 
  use of  these processes 
as luminosity monitor~\cite{mandy}.

\begin{figure}[htb]
\vspace*{6.2 cm}
\includegraphics{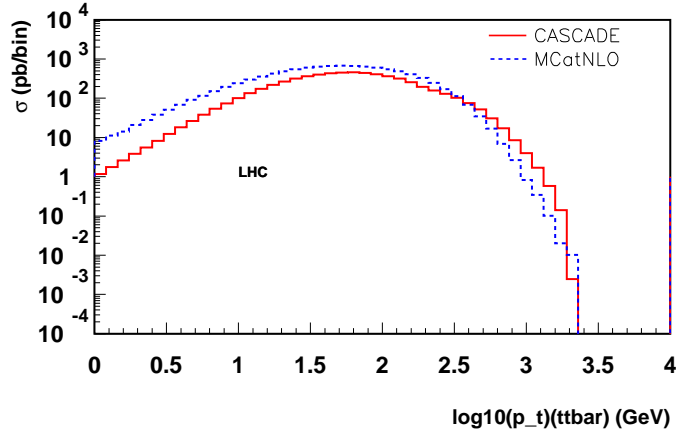}
\caption{Comparison of transverse momentum distribution of $t\bar{t}$ pairs calculated from
the k$_\perp$-shower  \cascade\ 
with the  NLO  calculation \mcatnlo\  at LHC energies.} 
\label{fig:ttbar}
\end{figure}

 The k$_\perp$-shower method discussed in this paper  
 can be used all the way up to   high  transferred-momentum scales. 
 As an illustration    in   Fig.~\ref{fig:ttbar}  we present  
 a numerical calculation for the transverse momentum spectrum of 
 top-antitop pair production at the   LHC. Small-$x$ effects are not large in this case. 
Rather,  this process  illustrates  how the shower    works in the region of 
  finite $x$  and large virtualities  on   the order of the top quark mass.   
 It is interesting to note that even at LHC energies the transverse momentum distribution 
of top quark pairs calculated from the k$_\perp$-shower is similar to what is obtained 
from a full 
NLO calculation (including parton showers, MC@NLO~\cite{mcatnlo}),  
with the  k$_\perp$-shower  giving a somewhat harder spectrum,  
 Fig.~\ref{fig:ttbar}.

We conclude by  observing  that 
using  off-shell matrix elements convoluted with unintegrated
parton distributions including explicit parton showering, many 
of the subleading effects are properly simulated both in ep collisions and at the LHC. 
 We have found that multi-jet  predictions   provide  
comparable results to  NLO calculations, where applicable, and 
 are much closer to the measurements in a region where significant 
higher order contributions are expected.
The results provide a strong motivation for systematic studies of 
k$_\perp$-dependent parton branching methods.

\vskip 0.6 cm

\newpage 

\noindent 
{\bf Acknowledgments}. 
We are grateful to S.~Chekanov for many discussions about the jet 
measurements. We gratefully acknowledge  useful   discussions 
with  
Z.~Nagy and A.~Savin.  Thanks to  the HERA machine crew and 
 experiments
for providing  precise and interesting  measurements of multi-jet 
correlations.
Part of this work was done while one of us (F.H.) was visiting DESY. 
We thank the DESY directorate for hospitality and support.

\appendix

\section{Fits to  the starting  pdfs }
\label{app:ktfits}

The branching equations   (\ref{uglurepr1}),(\ref{uglurepr2})   contain the  
starting gluon distribution ${\cal A}_0 ( x , \kt , Q_0 )$ at scale $Q_0$. 
\begin{figure}[htb]
\vspace{140mm}
\includegraphics{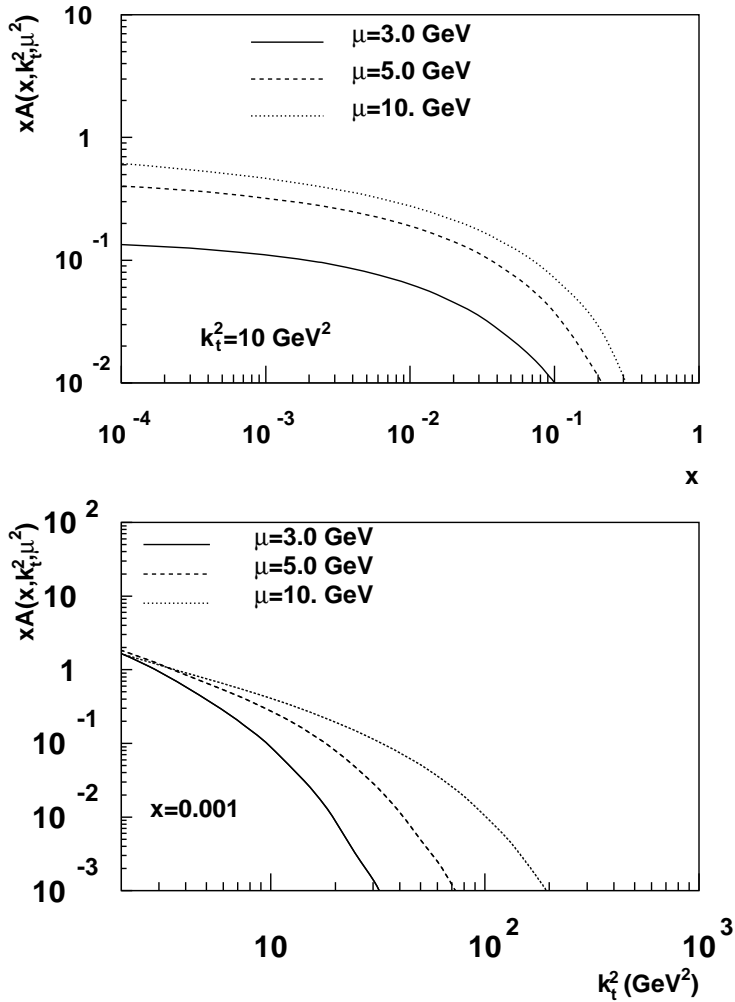}
\caption{(top) $x$-dependence  and 
(bottom) \kt-dependence of the unintegrated 
gluon distribution 
at different values of the evolution scale 
$\mu$.} 
\label{fig:ktglu}
\end{figure}
This is determined from fits to experimental data. In this appendix we  report  results 
for this distribution.

The starting  
 ${\cal A}_0 ( x , \kt , Q_0 )$     
at scale $Q_0$ is 
parameterized as~\cite{boand02,hj04} 
\begin{equation}
\label{ugluparam}
x {\cal A}_0  ( x , \kt, Q_0) = 
A \ x^{- B} \ (1 - x)^{C} \  \exp \left[ - (\kt - 
\lambda)^2 / \nu^2 \right] 
 \hspace*{0.3 cm}  .    
\end{equation}
The values of the 
parameters $A$, $B$, $C$, $\lambda$ and $\nu$    
in Eq.~(\ref{ugluparam}) are   determined 
from  data  fits~\cite{hj04,junghanss}.  
In the calculations  of the present  paper  we use the u-pdf set 
specified by the following parameter values: 
\begin{eqnarray}
\label{paramvalues}
 Q_0 &=& 1.1 \ {\rm{GeV}} \hspace*{0.3 cm} , \hspace*{0.6 cm}
       A = 0.4695 \hspace*{0.3 cm} , \hspace*{0.6 cm} 
       B = 0.025 \hspace*{0.3 cm} , 
\nonumber\\
       C &=& 4.0 \hspace*{0.3 cm} , \hspace*{0.6 cm} 
       \lambda = 1.5 \ {\rm{GeV}} \hspace*{0.3 cm} , \hspace*{0.6 cm} 
       \nu = (1.5/\sqrt{2}) \ {\rm{GeV}} 
 \hspace*{0.3 cm}  .    
\end{eqnarray}

In Fig.~\ref{fig:ktglu}   we plot    the $x$-dependence  and 
\kt-dependence of the 
resulting gluon distribution 
at different values of the evolution scale 
$\mu$.

\section{Time-like showering effects}
\label{sec_timelike}

The partons from the initial state cascade are allowed to develop a time-like
shower  in  \cascade\  2.0.2,  to be published in~\cite{deaketal}.  
 Full details    will be reported in this publication.  
 To give an idea of the effects,   we    include one  of the  results 
 in this appendix.  

\begin{figure}[htb]
\vspace*{11.1 cm}
\includegraphics{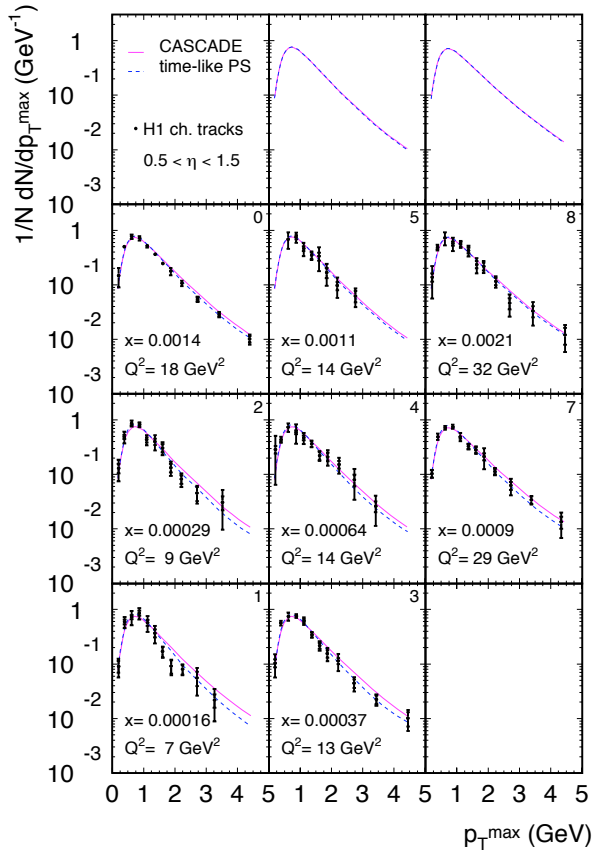}
\caption{The effect of including time-like  
showering on the ep charged-particle spectrum, along with 
the data~\protect\cite{h1charged}.} 
\label{fig:tlk}
\end{figure}

The
maximum scale for the time-like cascade is given by the transverse momentum of
the initial state gluon.   No additional constraints are applied to the
time-like shower. It is found that the  number of gluons after the initial state cascade with
time-like showering increases; however the effect on the  angular correlations 
  considered in Sec.~\ref{sec3} of this paper   is  
negligible (and smaller than the statistical error of the Monte Carlo  simulation).

In observables which are more sensitive to the time-like showering, like the 
charged-particle 
transverse momentum spectra (Fig.~\ref{fig:tlk}),  a  small effect coming from the
time-like showering can be observed,  and is of the same size as that 
obtained from Monte Carlo 
 event generators using the collinear parton-showering approach.


\begin{references}                        

\bibitem{bernhouches} 
     Z.~Bern  {\it et al.},  arXiv:0803.0494 [hep-ph], 
     in Proceedings of the Workshop 
     ``Physics at TeV Colliders" (Les Houches, 2007). 
\bibitem{mlmhoche}
         S.~H{\" o}che, F.~Krauss, 
         N.~Lavesson, L.~L{\" o}nnblad, 
         M.~Mangano, A.~Sch{\" a}licke and  S.~Schumann, 
         hep-ph/0602031;     N.~Lavesson and 
	  L.~L{\" o}nnblad,  arXiv:0712.2966 [hep-ph].  

\bibitem{alwalletal} 
     J.~Alwall  {\it et al.},  Eur.\ Phys.\ J.\  C {\bf 53} (2008) 473 
     [arXiv:0706.2569 [hep-ph]].  

\bibitem{heralhcproc}
     S.~Alekhin  {\it et al.}, 
     hep-ph/0601012,   hep-ph/0601013: Proceedings of 
     the Workshop ``HERA and the LHC",  
     eds. A.~De Roeck and      H.~Jung. 
	 
\bibitem{hj_rec}
 F.~Hautmann and H.~Jung,
  %``Recent results on unintegrated parton distributions,''
  arXiv:0712.0568 [hep-ph].  
  %%CITATION = ARXIV:0712.0568;%%   


\bibitem{herwref}
%\cite{Corcella:2000bw}
%\bibitem{Corcella:2000bw}
  G.~Corcella {\it et al.},
%``HERWIG 6: An event generator for hadron emission reactions with
%interfering gluons (including supersymmetric processes),''
  JHEP {\bf 0101} (2001) 010
  [arXiv:hep-ph/0011363]; 
%%CITATION = JHEPA,0101,010;%%
%\cite{Corcella:2002jc}
%\bibitem{Corcella:2002jc}
  G.~Corcella {\it et al.},
%``HERWIG 6.5 release note,''
  arXiv:hep-ph/0210213.
%%CITATION = HEP-PH/0210213;%%

\bibitem{pythref}
         T.~Sj{\" o}strand, S.~Mrenna and P.~Skands, JHEP 
         {\bf 0605} (2006) 026. 


\bibitem{mw92}
     G.~Marchesini and B.R.~Webber,
     Nucl.\ Phys.\ {\bf B386} (1992) 215.

\bibitem{skewang}
     M.~Ciafaloni, Nucl.\ Phys.\ {\bf B296} (1988)  49.

\bibitem{hef} 
     S.~Catani, M.~Ciafaloni and F.~Hautmann, Phys.\  Lett.\  
     B {\bf 242}  (1990) 97;   
     Nucl.\ Phys.\  B{\bf 366} (1991) 135; 
  Phys.\ Lett.\  B {\bf 307}  (1993) 147. 

\bibitem{boand02}
     B.~Andersson {\it et al.},   Eur.\ Phys.\ J.\  C {\bf 25} (2002) 77 
     [arXiv:hep-ph/0204115]. 

\bibitem{junghgs}
 H.~Jung, 
%``The CCFM Monte Carlo generator CASCADE,''
Comput.\ Phys.\ Commun.\  {\bf 143} (2002) 100
[arXiv:hep-ph/0109102]; 
%%CITATION = CPHCB,143,100;%%
H.~Jung and G.P.~Salam,
 %``Hadronic final state predictions from CCFM: The hadron-level Monte  Carlo
 %generator CASCADE,''
 Eur.\ Phys.\ J.\  C {\bf 19} (2001) 351
 [arXiv:hep-ph/0012143].   
 %%CITATION = EPHJA,C19,351;  

\bibitem{lonn}
    L.~L{\" o}nnblad and M.~Sj{\" o}dahl,  
    JHEP {\bf 0505} (2005) 038; 
    JHEP {\bf 0402} (2004) 042;  
    G.~Gustafson, L.~L{\" o}nnblad and G.~Miu, 
    JHEP {\bf 0209} (2002) 005.

\bibitem{golec-mc}
      K.~Golec-Biernat, S.~Jadach, W.~Placzek, 
      P.~Stephens and  M.~Skrzypek,       hep-ph/0703317.

 
\bibitem{krauss-bfkl}
      S.~H{\" o}che, F.~Krauss and T.~Teubner, 
      arXiv:0705.4577 [hep-ph]. 


   
     
\bibitem{zeus1931}
S.~Chekanov {\it et al.}  [ZEUS Collaboration],
%``Multijet production at low x(Bj) in deep inelastic scattering at HERA,'' 
Nucl.\ Phys.\  B {\bf 786} (2007) 152
  [arXiv:0705.1931 [hep-ex]].
%%CITATION = NUPHA,B786,152;%%



\bibitem{d02005}
       V.M.~Abazov  {\it et al.}  [D0 Collaboration],
       Phys.\ Rev.\ Lett.\ {\bf 94}  (2005) 221801
       [arXiv:hep-ex/0409040].


\bibitem{albrow}
        M.G.~Albrow   {\it et al.} [TeV4LHC QCD Working Group],
        arXiv:hep-ph/0610012.

\bibitem{aktas_h104}
    A.~Aktas {\it et al.}  [H1 Collaboration],
    Eur.\ Phys.\ J.\  C {\bf 33} (2004) 477 
    [arXiv:hep-ex/0310019].   

\bibitem{hansson06}
     M.~Hansson [H1 Collaboration], in Proceedings of the 
     14th International Workshop on Deep Inelastic 
     Scattering {\small DIS2006} 
     (Tsukuba, April 2006), p.~539.      

\bibitem{nagy}
%\cite{Nagy:2001xb}
%\bibitem{Nagy:2001xb}
Z.~Nagy and Z.~Trocsanyi,
%``Multi-jet cross sections in deep inelastic scattering at  next-to-leading
%order,''
Phys.\ Rev.\ Lett.\  {\bf 87} (2001) 082001 
[arXiv:hep-ph/0104315].
%%CITATION = PRLTA,87,082001;%%
%        Z.~Nagy and Z.~Trocsany, 
%        Phys.\ Rev.\ Lett.\ {\bf 87} (2001) 082001. 



\bibitem{h101} 
%\cite{Adloff:2000qj}
%\bibitem{Adloff:2000qj}
C.~Adloff {\it et al.}  [H1 Collaboration],
%``Measurement of neutral and charged current cross sections in electron
%proton collisions at high Q**2,''
Eur.\ Phys.\ J.\  C {\bf 19} (2001) 269
[arXiv:hep-ex/0012052].
%%CITATION = EPHJA,C19,269;%%       
%H1 Collaboration (C. Adloff et al.), 
%       Eur. Phys. J. C{\bf 19} (2001) 289.



\bibitem{nonglobjet} 
%\cite{Delenda:2006nf}
%\bibitem{Delenda:2006nf}
Y.~Delenda, R.~Appleby, M.~Dasgupta and A.~Banfi,
%``On QCD resummation with k(t) clustering,''
JHEP {\bf 0612} (2006) 044
[arXiv:hep-ph/0610242].
%%CITATION = JHEPA,0612,044;%%
%Y.~Delenda, R.~Appleby, M.~Dasgupta and  A.~Banfi, 
%JHEP {\bf 0612} (2006) 044.

\bibitem{dasguban}
        A.~Banfi and M.~Dasgupta, JHEP {\bf 0401} (2004) 027. 

\bibitem{delenda}
        Y.~Delenda, arXiv:0706.2172; 
        A.~Banfi,  M.~Dasgupta and  Y.~Delenda, arXiv:0804.3786. 


\bibitem{jcc-lc08}
      J.C.~Collins, talk at the Light Cone 2008 Workshop, Mulhouse, July 2008  
          [arXiv:0808.2665 [hep-ph]]. 

  
\bibitem{collins01}
     J.C.\ Collins, hep-ph/0106126,  in 
   Proceedings of the Workshop {\small DIS01}   (Bologna, 2001). 

\bibitem{jeppe04}
   J.R.~Andersen  {\it et al.},   Eur.\ Phys.\ J.\  C {\bf 35} (2004) 67.  


\bibitem{radcortalk1} 
      M.~Ciafaloni, talk at {\footnotesize RADCOR2007}, Florence,     October 2007.

\bibitem{radcortalk2} 
      G.~Altarelli, talk at {\footnotesize RADCOR2007}, Florence,     October 2007  
       [arXiv:0802.0968 [hep-ph]]. 

\bibitem{hef94} 
         S.~Catani and F.~Hautmann,
         Nucl.\ Phys.\ {\bf B427} (1994) 475;   
     Phys.\ Lett.\  B {\bf 315}  (1993) 157.






\bibitem{lip97}
       L.N.~Lipatov, Phys.\  Rept.\  {\bf 286} (1997) 131. 



\bibitem{hj04}
    H.~Jung, Mod.\ Phys.\ Lett.\ A{\bf 19} (2004) 1.

\bibitem{vannee}
   M.~Buza, Y.~Matiounine, J.~Smith, 
   R.~Migneron and  W.L.~ van Neerven, Nucl.\ Phys.\ {\bf B472} (1996) 611; 
   S.~Riemersma, W.L.~van Neerven  and J.~Smith, Phys.\ Lett.\  
   B {\bf 347} (1995) 143; E.~Laenen, 
   S.~Riemersma, W.L.~van Neerven  and J.~Smith, Nucl.\ Phys.\ 
   {\bf B392} (1993) 162.

\bibitem{mochetal}
    S.~Moch, J.A.M.~Vermaseren and A.~Vogt, 
    Phys.\ Lett.\  B {\bf 606}  (2005) 123;  
    Nucl.\ Phys.\ {\bf B691} (2004) 129. 



\bibitem{jccfh00}
    J.C.~Collins and F.~Hautmann,
    Phys.\ Lett.\ B {\bf 472} (2000) 129. 


\bibitem{rogers}
      T.C.~Rogers,   arXiv:0807.2430 [hep-ph];  arXiv:0712.1195 [hep-ph], 
      in Proceedings of the    8th International Symposium on Radiative 
     Corrections {\footnotesize RADCOR2007} (Florence,     October 2007); 
     J.C.~Collins,  T.C.~Rogers and  A.M.~Stasto,  
     arXiv:0708.2833 [hep-ph]. 

\bibitem{dyproof} 
      J.C.~Collins, D.E.~Soper and G.~Sterman, 
      Nucl.\ Phys.\  B{\bf 308} (1988) 833.  

\bibitem{collins0708}
     J.C.~Collins, arXiv:0708.4410 [hep-ph]. 

\bibitem{vogel0708}      
     W.~Vogelsang and F.~Yuan,  Phys.\ Rev.\ D {\bf 76} 
     (2007) 094013. 

\bibitem{bomhmuld}     
     C.J.~Bomhof and P.J.~Mulders, Nucl.\ Phys.\  B{\bf 795} (2008) 409. 


\bibitem{manch}
   J.R.~Forshaw, A.~Kyrieleis and M.H.~Seymour, 
   JHEP  {\bf 0608} (2006)  059;  M.H.~Seymour, arXiv:0710.2733 [hep-ph]. 

\bibitem{fhfeb07}
     F.~Hautmann,   Phys.\ Lett.\  B {\bf  655} (2007) 26.

\bibitem{korchangle}
     G.P.~Korchemsky and  A.~Radyushkin, 
     Phys.\ Lett.\ B {\bf 279} (1992) 359, 
     G.P.~Korchemsky and  G.~Marchesini,  
     Phys.\ Lett.\ B {\bf 313} (1993) 433.
 
\bibitem{chered} 
        I.O.~Cherednikov and N.G.~Stefanis,  
    Nucl.\ Phys.\  B{\bf 802} (2008) 146;  Phys.\ Rev.\ D {\bf 77} 
     (2008) 094001; arXiv:0711.1278 [hep-ph]. 

  


\bibitem{collsud}
    J.C.\ Collins, in {\em Perturbative Quantum Chromodynamics},
    ed.~A.H.~Mueller, World Scientific 1989, p.~573. 

 


\bibitem{anselm08} 
     M.~Anselmino et al.,  arXiv:0807.0173 [hep-ph]; 
      M.~Anselmino et al., 
    Phys.\ Rev.\ D {\bf 71}   (2005) 074006. 

\bibitem{ceccopieri}
     F.\ Ceccopieri and L.\ Trentadue,   Phys.\ Lett.\ B {\bf 660} (2008) 43. 

\bibitem{muldetal}
       A.~Bacchetta, D.~Boer, M.~Diehl and  P.J.~Mulders, 
     arXiv:0803.0227 [hep-ph].    


\bibitem{koike}
     Y.~Koike, W.~Vogelsang and  F.~Yuan,    
     Phys.\ Lett.\ B {\bf 659} (2008) 878; and references therein. 


\bibitem{bochumgpd}
     S.~Meissner,   K.~Goeke,  A.~Metz  
     and M.~Schlegel,     arXiv:0805.3165 [hep-ph]; and references therein. 


\bibitem{jccfh01}
        J.C.~Collins and F.~Hautmann,
    JHEP {\bf 0103} (2001) 016; F.~Hautmann, hep-ph/0101006. 


\bibitem{jiyuan}
    X.~Ji, J.~Ma and F.~Yuan, Phys.\ Rev.\ D {\bf 71}
    (2005) 034005, 
    JHEP {\bf 0507} (2005) 020.  

\bibitem{manohstew}
     A.V.~Manohar and I.W.~Stewart,  
     Phys.\ Rev.\  D {\bf 76} (2007) 074002. 
     
\bibitem{leesterm} 
     C.~Lee and G.~Sterman, 
     Phys.\ Rev.\ D {\bf 75} (2007) 014022. 

\bibitem{idimeh}
     A.~Idilbi and T.~Mehen, 
     Phys.\ Rev.\ D {\bf 75} (2007) 114017. 

\bibitem{bauergen}
   C.W.~Bauer, F.J.~Tackmann and J.~Thaler, 
   arXiv:0801.4028; 
   arXiv:0801.4026.

\bibitem{chay} 
        J.~Chay, arXiv:0711.4295. 

\bibitem{trottetal}
     C.W.~Bauer,  S.P.~Fleming, C.~Lee and G.~Sterman, 
     arXiv:0801.4569 [hep-ph];    
     J.~Chiu, F.~Golf, R.~Kelley and  A.V.~Manohar, 
     Phys.\ Rev.\ D {\bf 77} (2008) 053004; 
     M.D.~Schwartz,  Phys.\ Rev.\ D {\bf 77} (2008) 014026;  
     M.~Trott, Phys.\ Rev.\ D {\bf 75} (2007) 054011.  


\bibitem{fhdistalk}
     F.~Hautmann, arXiv:0708.1319 [hep-ph]. 

\bibitem{deaketal} 
      M.~Deak, H.~Jung et al.,   contribution to HERA-LHC 2008 Proceedings; in progress.   



\bibitem{radcor} 
     F.~Hautmann and H.~Jung, arXiv:0804.1746 [hep-ph], 
     in Proceedings of the 
     8th International Symposium on Radiative 
     Corrections {\footnotesize RADCOR2007} (Florence,     October 2007). 



\bibitem{dasgqt}
        M.~Dasgupta and  Y.~Delenda,    JHEP {\bf 0608} (2006) 080.

\bibitem{baines}
       J.~Baines et al., 
       Report of the Heavy Quark Working Group, 
       arXiv:hep-ph/0601164; 
       P.~Nason et al., Report on 
       bottom production, in  ``SM physics 
       at the     LHC",  arXiv:hep-ph/0003142. 



\bibitem{mcatnlo}
  S.~Frixione, P.~Nason and B.~R.~Webber,
  %``Matching NLO QCD and parton showers in heavy flavour production,''
  JHEP {\bf 0308} (2003) 007
  [arXiv:hep-ph/0305252].
  %%CITATION = JHEPA,0308,007;%%

\bibitem{mlm93}
      M.L.~Mangano,    Nucl.\ Phys.\  B {\bf 405} (1993) 536. 


\bibitem{higgs02}
        F.~Hautmann,  Phys.\ Lett.\ B {\bf 535} (2002) 159.


\bibitem{vogelhiggs}
        A.~Kulesza, G.~Sterman and W.~Vogelsang,
        Phys.\ Rev.\ D {\bf 69}    (2004) 014012.

\bibitem{olness}
      F.I.~Olness, talk at HERA-LHC Workshop, CERN, May 2008. 

\bibitem{cpyuan1} 
      S.~Berge, P.M.~Nadolsky,  F.I.~Olness and C.P.~Yuan, hep-ph/0508215; 
      P.M.~Nadolsky, N.~Kidonakis, F.I.~Olness and C.P.~Yuan, 
      Phys.\ Rev.\  D {\bf 67}  (2003) 074015. 

\bibitem{mandy} 
      A.M.~Cooper-Sarkar,  arXiv:0707.1593 [hep-ph].




\bibitem{junghanss}
%\cite{Hansson:2007de}
%\bibitem{Hansson:2007de}
M.~Hansson and H.~Jung,
%``Towards precision determination of uPDFs,''
arXiv:0707.4276 [hep-ph].
%%CITATION = ARXIV:0707.4276;%% 
%        M.~Hansson and H.~Jung, arXiv:0707.4276



\bibitem{h1charged} 
   C.~Adloff {\it et al.}  [H1 Collaboration],   Nucl.\ Phys.\  B {\bf 485} (1997) 3    
   [arXiv:hep-ex/9610006]. 




\end{references}
\end{document}